\documentclass[english,13pt]{scrartcl}
\usepackage[pages=all, color=black, position={current page.south}, placement=bottom, scale=1, opacity=1, vshift=5mm]{background}
\SetBgContents{}      
\usepackage[margin=1in]{geometry} 

\usepackage{amsmath}
\usepackage{amsthm}
\usepackage{amssymb}

\usepackage[utf8]{inputenc}
\usepackage{hyperref}
\hypersetup{
	unicode,
	pdfauthor={Christopher Hauer},
	pdftitle={Detection and Classification of
Cetacean Echolocation Clicks using
Image-based Object Detection
Methods applied to Advanced
Wavelet-based Transformations},
	pdfproducer={LaTeX},
}

\usepackage[english]{babel}

\theoremstyle{plain}

\theoremstyle{definition}

\usepackage{graphicx, color}
\graphicspath{{fig/}}

\usepackage{algorithm, algpseudocode} 
\usepackage{mathrsfs} 

\usepackage{lipsum}
\usepackage{subcaption}
\usepackage{setspace}
\usepackage[export]{adjustbox}
\usepackage[singlelinecheck=off]{caption}
\begin{document}
\begin{titlepage}
\begin{center}
    \begin{figure}[ht]
        \centering
        \includegraphics[height=2cm]{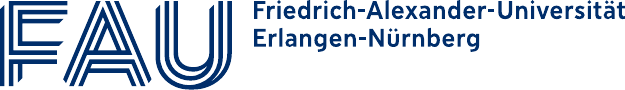}
    \end{figure}
\end{center}

\begin{center}

    \begin{spacing}{1.7}
    {
    \fontsize{22}{26} \selectfont \textbf{Detection and Classification of
Cetacean Echolocation Clicks using
Image-based Object Detection
Methods applied to Advanced
Wavelet-based Transformations}}\\[5mm]
    \end{spacing}
    \vspace{1cm}
    {\fontsize{18}{22} \selectfont Master Thesis}
    \vspace{4cm}
    
    \begin{tabular}{ll}

        Handed in by:       & Christopher Hauer   
            \\\\
        Matriculation number: & 23110234
        \\\\
        First supervisor:      & Alexander Barnhill    \\\\
        Second supervisor: & Prof. Dr.-ing Andreas Maier \\\\
        Advisor:    & Assistant Prof. Dr. Heike Vester   \\\\	
    \end{tabular}
\end{center}

    \centering
    Lehrstuhl für Mustererkennung (LME) | Martensstr. 3 | 91058 Erlangen | cs5-info@lists.fau.de

\end{titlepage}
\section*{Declaration of Originality}
I, Christopher Hauer, student registration number: 23110234, hereby confirm that I completed the submitted work independently and without the unauthorized assistance of third parties and without the use of undisclosed and, in particular, unauthorized aids. This work has not been previously submitted in its current form or in a similar form to any other examination authorities and has not been accepted as part of an examination by any other examination authority.

Where the wording has been taken from other people’s work or ideas, this has been properly acknowledged and referenced. This also applies to drawings, sketches, diagrams and sources from the Internet.

In particular, I am aware that the use of artificial intelligence is forbidden unless its use as an aid has been expressly permitted by the examiner. This applies in particular to chatbots (especially ChatGPT) and such programs in general that can complete the tasks of the examination or parts thereof on my behalf.

Furthermore, I am aware that working with others in one room or by means of social media represents the unauthorized assistance of third parties within the above meaning, if group work is not expressly permitted. Each exchange of information with others during the examination, with the exception of examiners and invigilators, about the structure or contents of the examination or any other information such as sources is not permitted. The same applies to attempts to do so.

Any infringements of the above rules constitute fraud or attempted fraud and shall lead to the examination being graded “fail” (“nicht bestanden”).
\newpage

\sloppy
\hyphenpenalty 10000
\exhyphenpenalty 10000
\section*{Abstract}
A challenge in marine bioacoustic analysis is the detection of animal signals, like calls, whistles and clicks, for behavioral studies. Manual labeling is too time-consuming to process sufficient data to get reasonable results. Thus, an automatic solution to overcome the time-consuming data analysis is necessary. Basic mathematical models can detect events in simple environments, but they struggle with complex scenarios, like differentiating signals with a low signal-to-noise ratio or distinguishing clicks from echoes. Deep Learning Neural Networks, such as ANIMAL-SPOT, are better suited for such tasks. DNNs process audio signals as image representations, often using spectrograms created by Short-Time Fourier Transform. However, spectrograms have limitations due to the uncertainty principle, which creates a tradeoff between time and frequency resolution. Alternatives like the wavelet, which provides better time resolution for high frequencies and improved frequency resolution for low frequencies, may offer advantages for feature extraction in complex bioacoustic environments.
This thesis shows the efficacy of CLICK-SPOT on Norwegian Killer whale underwater recordings provided by the cetacean biologist Dr. Vester. \\\\
Keywords: Bioacoustics, Deep Learning, Wavelet Transformation
\tableofcontents

\section{Introduction}
The field of bioacoustics is not just concerned with animal vocalization signal production, but also with their behavioral context and intra-species communication. 
To gather insight into the social group behavior and dynamics of animals, bioacousticians use data analysis on collected audio data from observed target animals. This analysis helps create hypotheses based on observed behavior, correlates reoccurrences of acoustic signals, and predicts the semantic meaning of animal communication.
To that regard, one successful starting strategy is trying to map reoccurring signals towards a specific animal behavior \cite{BirdGuide},
for example, to identify mating calls, warning calls, contact calls, begging calls, alarm calls, search calls and more \cite{KONDO09}. 
While this process has been successful in many species, such as many avian species \cite{MARLER96, Suzuki21} and mammals \cite{Seyfarth03}, the method fails in deciphering the contextual meaning from more complex communications \cite{Freeberg12}.
One such animal species is the charismatic toothed killer whale (\textit{Orcinus orca}) \cite{Ford:1987,Ford:1989,Ford:1991,Ford:2000,Towers:2015,Towers:2019,Ness:2013,Filatova:2007,Filatova:2015,Baird:2001,Ivkovich:2010}.
A killer whale vocal repertoire has been catalogued for killer whale populations in different regions around the world, such as the resident population in Canada \cite{Ford:1987,Ford:1989,Ford:1991,Ford:2000,Filatova:2007,Filatova:2015, OrcaCallCatalog}, Iceland \cite{Selbmann23} and Norway \cite{Vester17}.
For similar animals, such as the bottlenose dolphin, specific call types like signature whistles \cite{JANIK98} were found. While similar hypotheses were put forward on the observation of captive killer whales \cite{Parijs04}, so far, single call types have not yet been assigned to a specific animal behavior \cite{Wei07}.
The process of deciphering killer whale vocalizations and assigning them to animal behavior is particularly challenging because killer whales are per se hard to observe. The marine environment complicates efforts to track their behavior, and the migratory patterns make long-term study virtually impossible.
To gain deeper insight into killer whale communication, different approaches to semantic communication, such as call phonetics and turn-taking dialogs and conversations could help to determine context-dependent vocalizations.    

\subsection{Killer Whale Vocal Repertoire}
The vocal repertoire for killer whales is region-dependent, as such there are differences between the Canadian call catalog \cite{OrcaCallCatalog2}, the Icelandic call catalog \cite{Selbmann23, Giovannini25} and the Norwegian call catalog \cite{Vester17}. Yet the underlying structure of the catalog is similar. The killer whale vocalizations are categorized as calls, whistles and clicks. 
\subsubsection{Killer Whale Calls}
Killer whales use stereotyped calls for both short-range and long-range communication \cite{Miller06}. Low-frequency long-range calls can have a possible range of 16 kilometers and are more commonly recorded during feeding and foraging. Short-range calls are recorded during socializing and resting periods and can have a range of up to 9 kilometers. 
\subsubsection{Killer Whale Whistles}
Whistles are recorded more often during social interaction. As stated, it is suspected that killer whales use signature whistles for 
group recognition or kin recognition \cite{Parijs04}. However, proving the hypothesis is difficult due to the killer whales' elusive underwater lifestyle.
\subsubsection{Killer Whale Clicks}
In the ocean, sunlight decreases rapidly with depth and the  visibility strongly reduces depending on water quality and depth \cite{NOAA_Sunlight}. 
Given these circumstances, toothed whales, such as the killer whale, rely on echolocation for hunting and navigation.
Echolocation works just like a biological sonar \cite{Seaworld_Orca}, the killer whale produces a powerful impulse using its phonic lips in the nasal sacs. 
The impulse is guided and emitted through the melon, a round-shaped organ in the killer whale's forehead which consists of fat. 
The melon acts like an acoustical lens to focus the clicks into a directional beam. 
The returning echo is captured by the fat-filled cavities in the lower jar bone and received in the auditory bulla. 
During navigation or hunting, the killer whale can emit hundreds of clicks in rapid succession, known as a click train or burst, to target specific areas of interest or gather continuous information about their prey \cite{Holt19}. However, it is also suspected that clicks could play a role in communication.

\subsection{Current Situation} 
\begin{figure}[!htb] \centering
\begin{subfigure}[t]{0.30\textwidth}
  \includegraphics[width=\linewidth,valign=t]{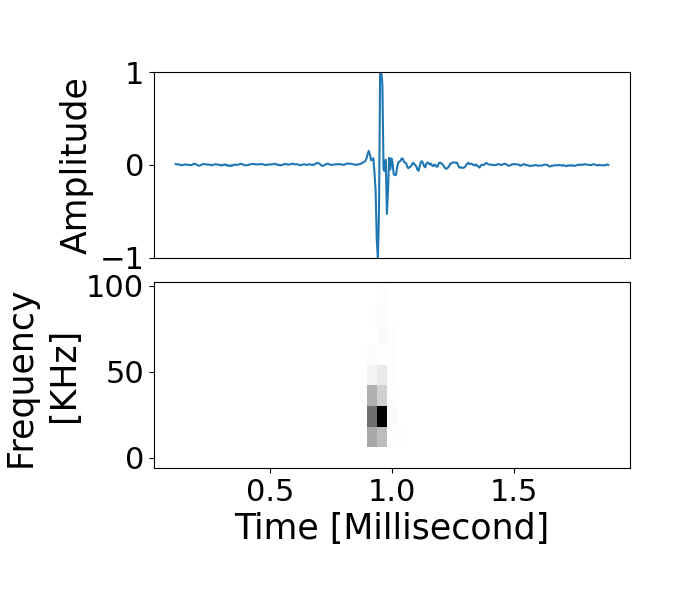} 
  \caption{Waveform and spectrogram of a high SNR click}
\end{subfigure}\hfil 
\begin{subfigure}[t]{0.30\textwidth}
  \includegraphics[width=\linewidth,valign=t]{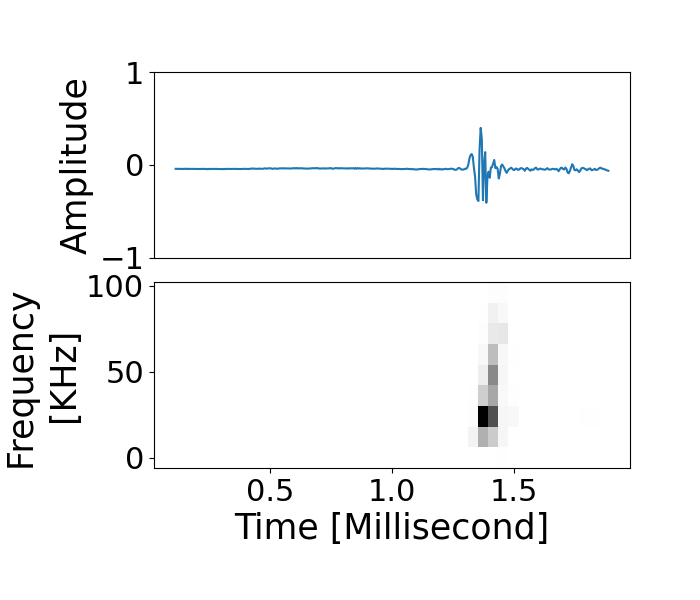}
  \caption{Waveform and spectrogram of a medium SNR click}
\end{subfigure}\hfil 
\begin{subfigure}[t]{0.30\textwidth}
  \includegraphics[width=\linewidth,valign=t]{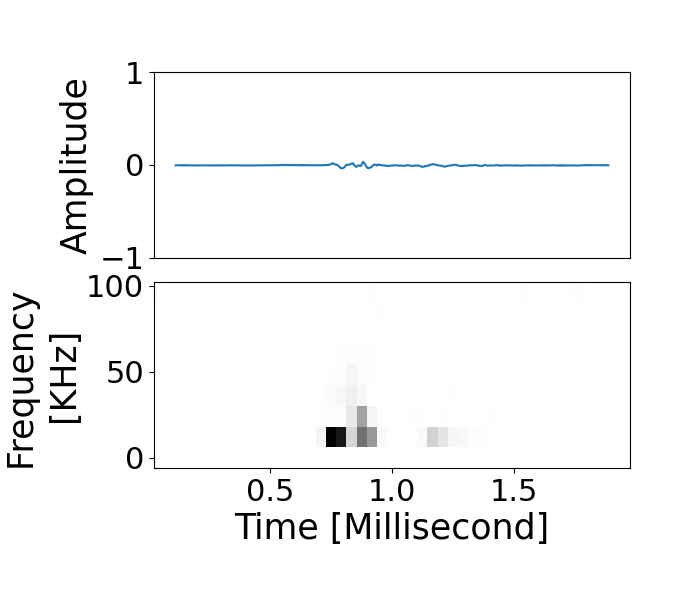}
  \caption{Waveform and spectrogram of a low SNR click}
\end{subfigure}

\medskip

\begin{subfigure}[t]{0.30\textwidth}
  \includegraphics[width=\linewidth,valign=t]{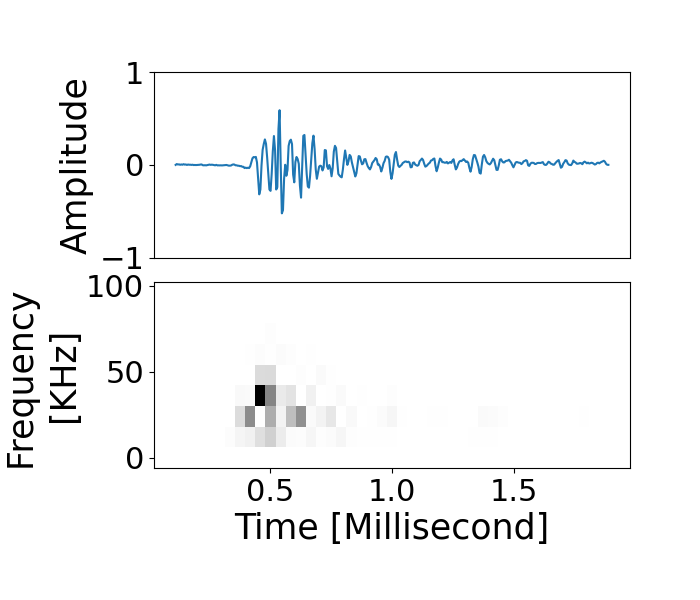}
  \caption{Waveform and spectrogram of a high SNR echo}
\end{subfigure}\hfil 
\begin{subfigure}[t]{0.30\textwidth}
  \includegraphics[width=\linewidth,valign=t]{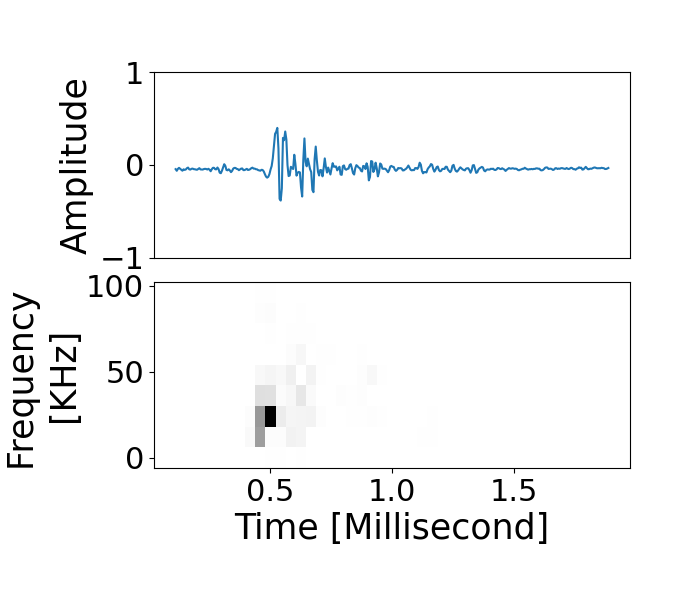}
  \caption{Waveform and spectrogram of a medium SNR echo}
\end{subfigure}\hfil 
\begin{subfigure}[t]{0.30\textwidth}
  \includegraphics[width=\linewidth,valign=t]{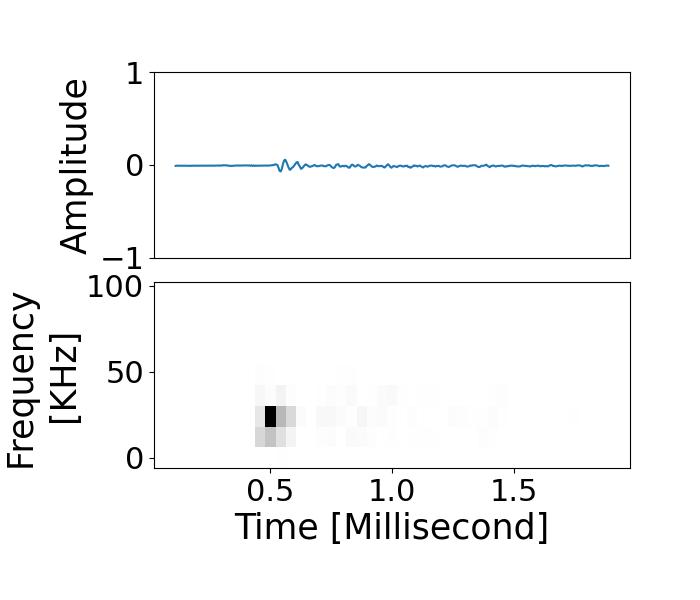}
  \caption{Waveform and spectrogram of a low SNR echo}
\end{subfigure}
\caption{Six example images of the waveform and spectrogram of different clicks and echoes in isolation. The differing SNR values can happen due to the distance of the emitted click to the hydrophone and the directionality of clicks. The spectrogram was generated with a segment size of 16 samples and a hop of 8 samples. The number of samples are very low when compared to the more commonly used 1024 segment size and 512 hop. This is necessary due to the small window size of only 384 samples (2 milliseconds using 192kHz). This explanation also holds for other pictures.} 
\label{fig:exampleclicks}
\end{figure}

To better understand the killer whale's use of clicks, they should be analyzed in context to determine whether they serve a communicative purpose.
To do this, the clicks need to be detected and annotated, which can pose numerous problems. 
The most common problem is that both the killer whale clicks and the returning echoes are recorded. As echoes are reflected clicks from surfaces, such as the ocean floor and the water surface, they share many characteristics with their original click. Since echoes are dependent on the environment, they should be differentiated from the clicks for better quantifiable measurements. 
Yet, the differentiation between clicks and echoes is difficult, as the versatility of clicks in different environments and the similarity between clicks and echoes makes it difficult to differentiate the two in isolation. 
The six images in Figure \ref{fig:exampleclicks} show some examples of the different clicks and echoes encountered in the field from different signal-to-noise ratio (SNR) environments.

\subsubsection{Dialog with Dr. Heike Vester}
\begin{figure}[!htb] \centering
\begin{subfigure}[t]{0.30\textwidth}
  \includegraphics[width=\linewidth,valign=t]{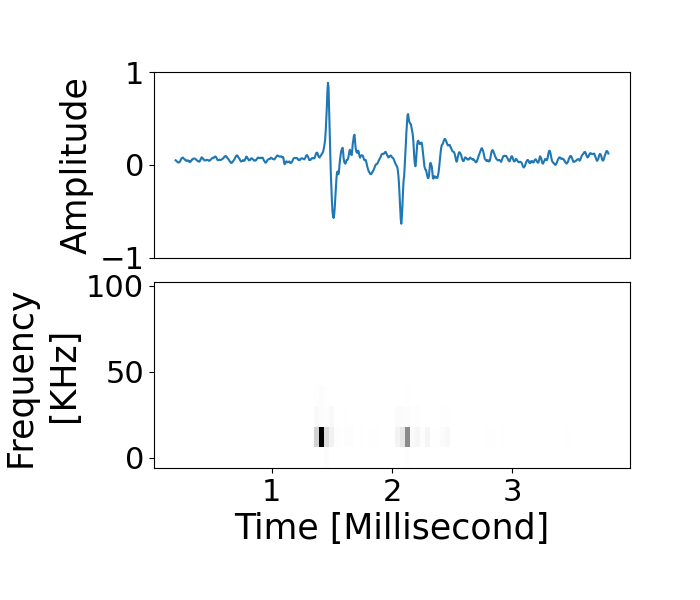}
  \caption{Waveform and spectrogram of a common click and echo pair}
\end{subfigure}\hfil 
\begin{subfigure}[t]{0.30\textwidth}
  \includegraphics[width=\linewidth,valign=t]{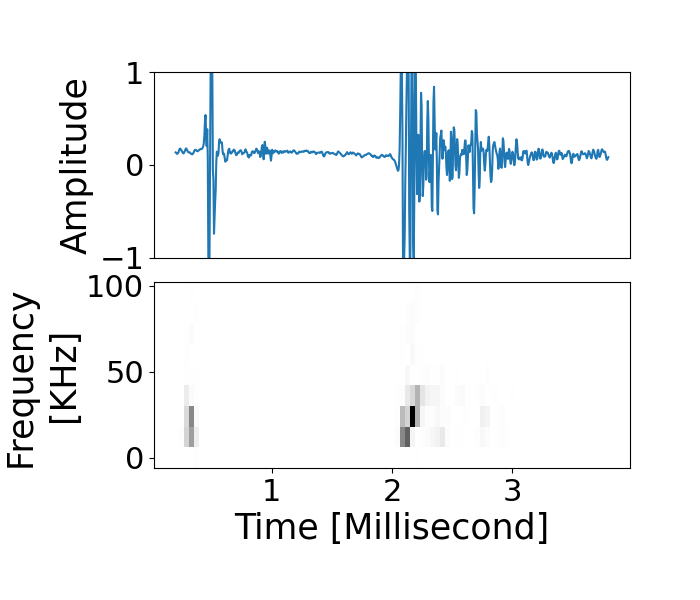}
  \caption{Waveform and spectrogram of a pair with the same starting amplitude}
\end{subfigure}\hfil 
\begin{subfigure}[t]{0.30\textwidth}
  \includegraphics[width=\linewidth,valign=t]{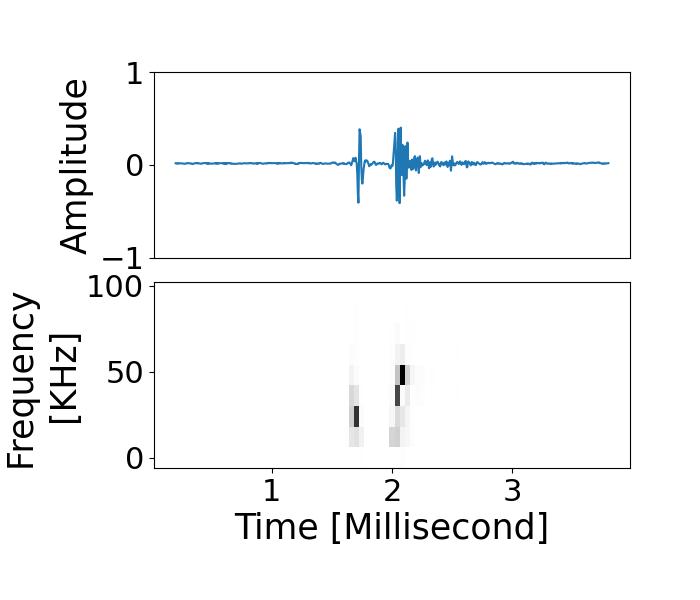}
  \caption{Waveform and spectrogram of a pair with inverted starting amplitude}
\end{subfigure}
\caption{Three example images of click and echo pairs. Image (a) shows a usual click and echo pair, with a positive starting amplitude click and a negative starting amplitude echo. Due to a weak initial phase, the echo in image (b) appears to have a positive starting amplitude. In Image (c), due to the strong impulse compared to the initial amplitude, it looks like the click starts with a negative amplitude and echo with a positive amplitude. }
\label{fig:examplePairs}
\end{figure}

\begin{figure}[!htb] \centering
\includegraphics[width=\linewidth,valign=t]{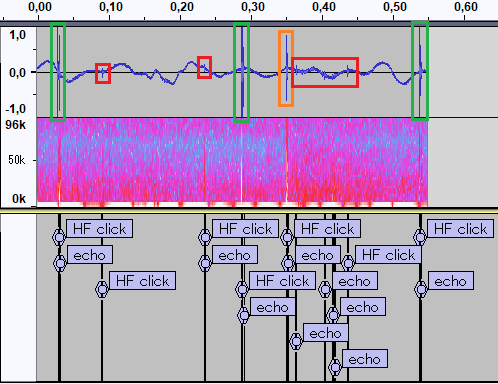}
\caption{An example of the waveform, spectrogram and labels in Audacity. The equidistant click echo pairs marked within green boxes are part of a burst. The click and echo pairs marked with orange boxes are also easy to distinguish, but not part of the same burst as the ones marked with green. The events marked with red boxes are labeled, but difficult to identify and potentially wrong.}
\label{fig:sourroundingExample}
\end{figure}

\begin{figure}[htb] \centering
\begin{subfigure}[t]{0.30\textwidth}
  \includegraphics[width=\linewidth,valign=t]{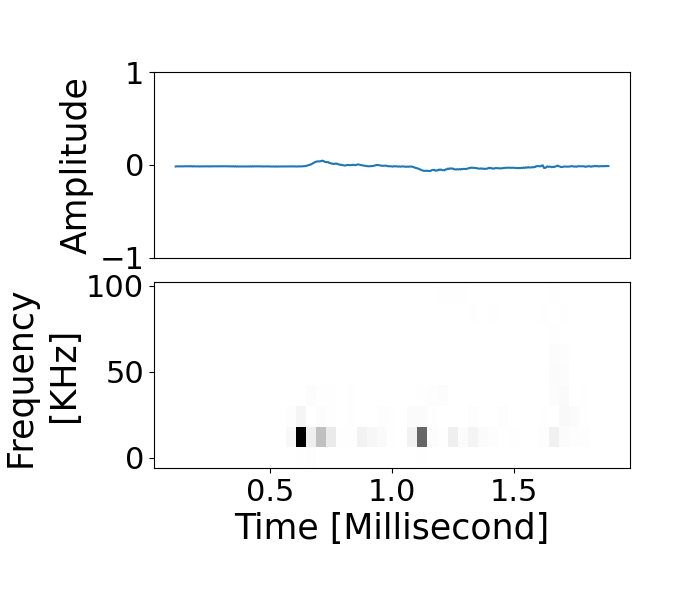}
  \caption{Low-Frequency (LF) click}
\end{subfigure}\hfil 
\begin{subfigure}[t]{0.30\textwidth}
  \includegraphics[width=\linewidth,valign=t]{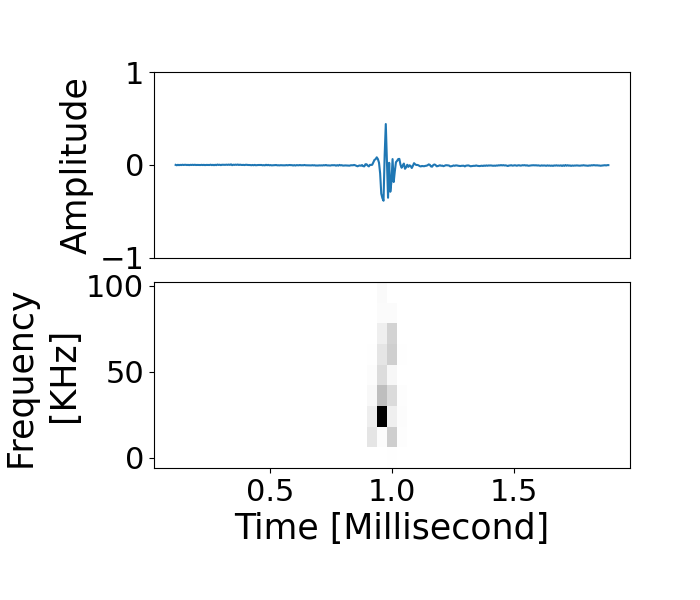}
  \caption{High-Frequency (HF) click}
\end{subfigure}\hfil 
\begin{subfigure}[t]{0.30\textwidth}
  \includegraphics[width=\linewidth,valign=t]{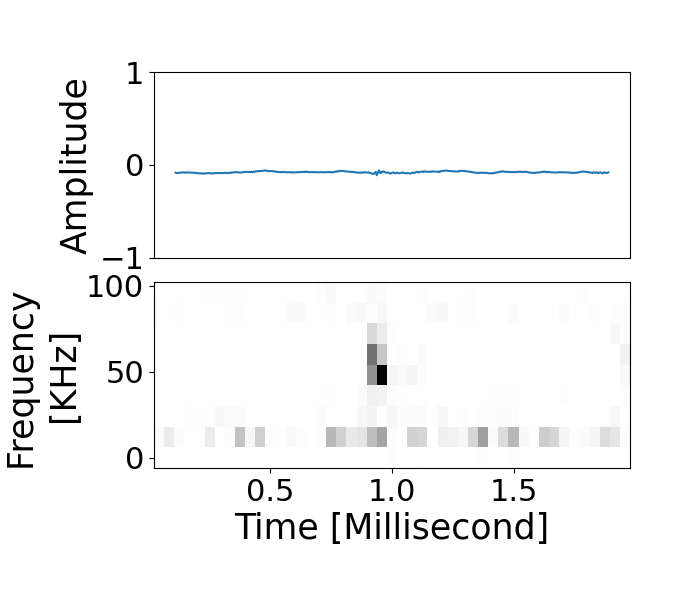}
  \caption{Ultrasonic-Frequency (US) \\click}
\end{subfigure}
\caption{Example of a low-frequency click [image (a)], a high-frequency click [image (b)] and an ultrasonic-frequency click [image (c)].}
\label{fig:examplelabels}
\end{figure}

To understand the problem with detecting and annotating clicks, Dr. Heike Vester, a biologist who specializes in the social behavior and bioacoustic of marine mammals, was interviewed to share her experiences with detecting and hand labeling clicks from the Norwegian killer whale population.
Stationed in Bod\o, Norway, she is the founder of Ocean Sounds e.V. \cite{Oceansounds} and provided both the task and the data for this master thesis. Following is a reproduced summary of the interview with Dr. Vester at the Ocean Sounds studio.
\begin{enumerate}
    \item \textbf{How is the data collected?}\\
    The data are only collected when the orca is visible and identified via PhotoID. A matriline can be identified by its members, which can be identified by their saddle patch and fin. A single hydrophone with a sampling rate of 192 kHz is lowered from the boat to record the animals. In addition, the animals' behavior during the recording is noted \cite{Vester17}. 
    
    \item \textbf{Why are clicks so numerous and important?}\\
    Killer whales primarily use echolocation to both orient themselves and to find and track prey during a hunt. But clicks are also emitted numerously during social interactions. Observing more animals, like during social interaction, will gain more clicks than observing a single group with fewer animals, of course. But the purpose of clicks during social interactions, where the animals can visually see each other, remains a mystery. A quantifiable analysis of clicks could help to decipher the meaning and usage of clicks for killer whales in situations where it might not be necessarily used for echolocation alone. 
    
    \item \textbf{How are the data hand labeled?} \\
    The data is hand labeled using Audacity \cite{Audacity}. By searching through the waveform of a signal to find high-frequency Dirac-like impulses or energy-rich impulses. These impulses, which span through the entire bandwidth, can also be found in the spectrogram. The events are marked with text annotations using a separate label track.
    
    \item \textbf{How can one tell clicks and echoes apart from each other?}\\
    Clicks and echoes are differentiated through the context of the event. Looking at an individual impulse usually does not provide enough information. The killer whales emit bursts of clicks to generate an image of their surrounding. These clicks are generated in somewhat equidistant time intervals. Of course, they are not exactly equidistant, these bursts can speed up and slow down in event frequency when the killer whale zooms into an area of interest. Strong and obvious clicks and echoes can be told apart by their different phases, a click usually starts with a positive amplitude, an echo with an inverted negative phase, but that is not always the case as can be seen in Figure \ref{fig:examplePairs}.
    The robust way of distinguishing clicks and echoes is by comparing the impulse with their neighboring patterns. For instance, if the time intervals are roughly equidistant, like in a burst, or if the click is significantly more intense than the echo, distinguishing between them becomes easy, as can be seen in Figure \ref{fig:sourroundingExample}. If a click is weaker or does not follow the burst structure, then it is more difficult to distinguish, in those cases, it takes some experience and intuition to differentiate between clicks and echoes. 

    \item \textbf{What types of labels are used?} \\
    The obvious clicks are divided into three subtypes based on their peak (most energy) frequency range in the spectrogram. 
    If a click has a peak below 5 kHz, it is labeled as a low-frequency (LF) click. If it has a peak between 5 kHz and 40 kHz, it is labeled as a high-frequency (HF) click. In the case of a peak above 40 kHz, the click is labeled as an ultrasonic-frequency (US) click. Figure \ref{fig:examplelabels} shows the three different labeled clicks. In comparison to LF and HF clicks, US clicks are rare. Sometimes it is difficult to determine the max energy in weaker clicks, which do not have a lot of energy and are barely visible in the spectrogram and time signal. These weak clicks are sometimes annotated with a suggestion on what the frequency could probably be, like weak LF clicks. Since the echoes are less critical to the analysis, they are not differentiated by peak frequency and are simply labeled as echoes.

    \item \textbf{Are there problems with the hand labeling?}\\
    The biggest problem is time. During a burst, a killer whale can emit more than a hundred clicks in less than a second. These clicks can have multiple echoes, so it is possible to have hundreds of events in a single second. Assuming it takes an experienced bioacoustician ten seconds to locate, mark, and compare an event with its surrounding context to create a label, the task becomes significantly more time-consuming when there are multiple events to label. For instance, if a burst contains 150 clicks and 210 echoes within a single second, the bioacoustician would need approximately one hour to label just that one second of data.
    Of course, not every second of the data material contains a burst. However, it took twelve hours to label a single minute of data material. This is just not sustainable for a good data analysis, even with the help of a team. 

    \item \textbf{For a start, would an event detector help in speeding up the analysis?}\\
    If an event detector can find and pre-mark the clicks and echoes, then that could help, but it would not solve the actual problem. The labeling of the event still takes a significantly long time. Even if the annotation process could be accelerated to be twice as fast, taking six hours to label just one minute of data is still unsustainable for thorough data analysis. This task demands a fully automated solution.

    \item \textbf{Given the problem, there is a high probability that the machine would not be as reliable as an experienced biologist. What would be a good compromise?}\\ 
    The first goal is to quantify the click occurrences and correlate them to animal behavior, so that we predict their behavior in the future without the need to see the animals, e.g. feeding events versus socializing or traveling.
    To achieve this, the clicks need to be compared with the animal's activity and behavior. However, the process does not need to be flawless from the outset. For instance, if the machine is accurate in 90\% of the cases, the results can still be used for further analysis.

    \item \textbf{Concerning true positive, false positive, true negative, false negative cases, what should be the focus?} \\
    The fully automated solution should maximize the overall accuracy. If a compromise exists between missing click edge cases and minimizing false positive findings, then the focus should be on minimizing the total number of errors between both false positive and false negative cases.

    \item \textbf{Human intuition is difficult to define, what features could be used for click and echo differentiation?} \\
    Assuming that the event detection is working, the inter-arrival-time between events, the energy density and max energy as well as the starting phase could be useful features for distinguishing clicks and echoes. These would need to be compared between multiple events to get the bigger picture. Of course, the maximum energy over the frequency is needed for the LF, HF and US differentiation.
\end{enumerate}

\subsection{Related Work}
Bioacoustic click event detection is not a new field of study. There are many bioacoustic toolkits which provide click detectors for passive acoustic monitoring based on different approaches.
SEDNA \cite{SEDNA}, developed by the Bioacoustic Research Program at Cornell University and the Lab of Ornithology, provides a bioacoustic click detector based on MATLAB tools for the use of environmental impact assessment. 
The work of C. Gervaise et al. \cite{GERVAISE2010} uses a click detector based on the kurtosis of a signal to research the psychological impact of noise pollution. 
To the best of the author's knowledge, the most widely used open-source click detector would be the PAMGuard click detector \cite{PamGUARD:2021}, developed by Jamie Macaulay, Doug Gillespie and Michael Oswald. 
The PAMGuard click detector is based on Java and uses energy over frequency thresholds to find clicks.
Another approach for calls rather than clicks is ANIMAL-SPOT \cite{Bergler21-AAA}, a ResNet18-based Convolutional Neural Network (CNN) that performs detection segmentation using a sliding window approach. The ANIMAL-SPOT model has been a major influence on the development of this thesis.
The two works of Bermant et al. \cite{Bermant19, Bermant2022} provide machine learning tools to find and annotate clicks and codas of sperm whales. The first machine learning techniques were based on Convolutional Neural Networks (CNNs) to extract finer-scale details from cetacean spectrograms \cite{Bermant19}. The later work provided a self-supervised deep learning method which uses Noise Contrastive Estimation to find spectral changes \cite{Bermant2022}. 
The Listening-Lab Annotator developed by McEwen et al. \cite{MCEWEN24} utilizes a wavelet-based segmentation method that automatically extracts transient features. However, the approach still relies on human-in-the-loop intervention for classification. This work has significantly influenced the development of this thesis, serving as a key inspiration for the methodology employed here.

\subsection{Contribution}
Many of the bioacoustic click detectors in passive acoustic monitoring use threshold-based baseline methods, which do not perform well in environments with a low signal-to-noise (SNR) ratio or do not perform adequate click echo differentiation without the need for human correction. 
This thesis focuses on the detection and differentiation of killer whale clicks and echoes to develop an event detector capable of identifying both clicks and echoes, along with implementing a method for distinguishing between the two. To achieve this, several approaches were tested, utilizing waveform, spectrogram, and scalogram audio representations in conjunction with ANIMAL-SPOT and YOLO models. 
In this work, the tool-chain CLICK-SPOT is introduced. CLICK-SPOT is enhancing a YOLO model with a First order detection (FOD) post processor and a random forest context classifier to find, annotate and label click and echoes from an audio file of an underwater recording. CLICK-SPOT was developed in three stages, the first alpha version only consisted of the YOLO model and merged bounding boxes. The second beta version added the FOD post processor to enhance the bounding boxes. The final CLICK-SPOT version added a random forest context classifier to differentiate clicks from echoes. 
Although these tools are specifically designed for detecting killer whale clicks and echoes, many of the methods and models can be adapted for identifying target signals of other animal species as well.

\subsection{Overview of the next Chapters}
The following chapters are structured as follows. 
In the theory chapter, the underlying theory for ANIMAL-SPOT and YOLO are described and summarized. 
The methodology provides an overview of the ANIMAL-SPOT and YOLO models, as well as the context approach used in the experiments and results. 
The Data chapter provides a description of the data provided by Dr. Vester from Ocean Sounds, as well as its usage and preprocessing for ANIMAL-SPOT and YOLO training. 
The experiments chapter describes the experiments performed in this thesis and reports on the results achieved in the experiments. 
The discussion chapter explains what problems occurred during the experiments, what could be learned from the experiment, and what problems occurred during the experiments. 
The conclusion summarizes this thesis and gives a further outlook on what could be done in the future. 

\section{Theory}
This chapter will give a summary of all the theories necessary to describe the task and methodology used in this thesis. 
It will first start with the physical attributes of clicks, echoes and the surrounding noise in underwater bioacoustics before diving into the technical aspects of First Order Detection (FOD), ANIMAL-SPOT, YOLO and context-dependent analysis. 
\subsection{Sound Characteristics of Clicks, Echoes and other Noises}
Sound travels faster underwater (approximately $1500m/s$) than through the air (approximately $~343m/s$). 
While there are differences in travel time and distance attenuation between underwater and air acoustics, the underlying theory remains the same for both systems.
To understand the problems and challenges in the following chapters, the attributes of orca clicks and echoes as well as other noises have to be analyzed and explored first.

\subsubsection{Sound Characteristics of Clicks}
A click is a Dirac-like impulse, typically starting with a positive amplitude. 
As stated in the introduction, the killer whale emits these clicks, which usually feature low-frequency (LF) peaks between 20 Hz and 5 kHz, high-frequency (HF) peaks between 5 kHz and 40 kHz \cite{Holt08}, or ultrasonic-frequency (US) peaks over 40 kHz. 
US clicks are less commonly observed compared to LF and HF clicks. This may be due to the shorter transmission range of higher-frequency sounds underwater, coupled with the greater directionality of these clicks, making them less likely to be recorded. Alternatively, it could simply be that these clicks are used less frequently in vocalizations.
Examples of these clicks can be seen in Figure \ref{fig:examplelabels}. 
The energy intensity of a killer whale click is difficult to estimate \cite{Holt08} due to the directional beam forming and high variation.  
The killer whale can produce clicks with a strength similar to that of the sperm whale, ranging from 188 decibels (dB) to possibly as high as 230 dB. 
From measurements of the provided data, on average, a killer whale click lasts less than 1 millisecond, and the inter-arrival time between clicks in a burst can be shorter than 2 milliseconds. 
The click typically begins with a positive amplitude impulse, but due to the underlying electrical noise floor and electrical system noise, determining the starting amplitude becomes challenging for clicks emitted far away from the hydrophone. 
The noise can distort the signal and affect both the minimum and maximum energy spikes. 
As a result, the prominent energy spikes are, while still useful, not fully reliable indicators for distinguishing between clicks and echoes.

\subsubsection{Sound Characteristics of Echoes}
\begin{figure}[!htbp] \centering
\begin{subfigure}[t]{0.45\textwidth}
  \includegraphics[width=\linewidth]{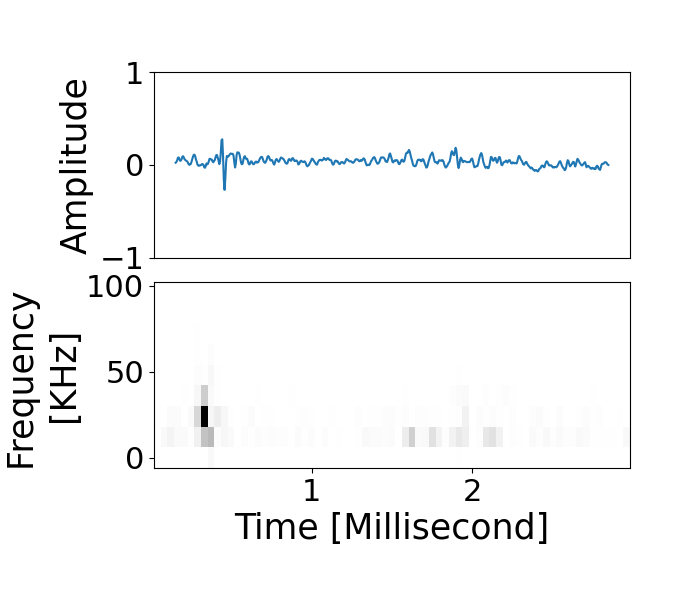}
  \caption{An example of a click and echo pair. The echo is more difficult to determine due to reverberations.}
  \label{fig:echoreverb}
\end{subfigure}\hfil 
\begin{subfigure}[t]{0.45\textwidth}
  \includegraphics[width=\linewidth]{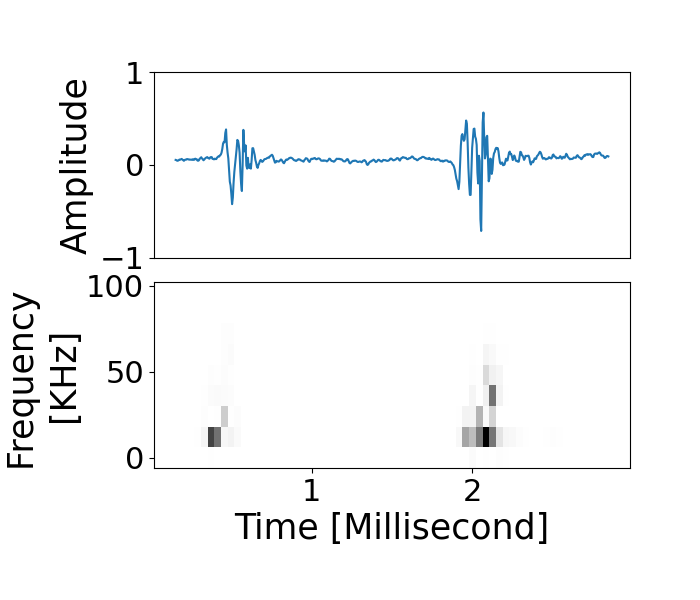}
  \caption{ An example of a click and echo pair. Due to the frequency shift of the echo and reverberation, the energy in certain bands can be higher than the original click, leading to a higher peak and/or average energy in the echo rather than the click.}
  \label{fig:echoestrongerclick}
\end{subfigure}\hfil 
\caption{Examples of a reverberated echo in image (a) and a frequency shifted reverberated echo in image (b). The echo in image (b) has a higher peak and average energy than the click. }
\label{fig:Noises}
\end{figure}

Echoes are bounced reflections of the click from a surface. From the provided data it can be seen that the echoes can arrive anywhere between less than a millisecond and 20 or more milliseconds dependent on the underwater travel time. On average, surface reflected echoes arrive 1.5 milliseconds after the click. Variations in click echo interarrival time can also be seen in Figure \ref{fig:examplePairs}. 
The echo exhibits a similar structure to the click but with an inverted starting amplitude. Both clicks and echoes experience attenuation as they travel through water, causing reverberation effects. The peak frequency of the echo can shift by 0.1 kHz to 2 kHz, influenced by factors such as the angle and properties of the reflective surface. These reverberations complicate the precise identification of the echo's start and end points, as can be seen in Figure \ref{fig:echoreverb}.

Interestingly, due to frequency shifts and potential overlap with underlying noise, the echo can have a higher peak and/or average energy than the click. One example of such an occurrence of a click and echo pair with a stronger echo can be seen in Figure \ref{fig:echoestrongerclick}.
That means that the maximum and average energy alone are unreliable indicators for differentiating between click and echo.

\subsubsection{Click and Echo Differentiation} 
When seen as a pair, differences from attenuation and reverberation are apparent, as can be seen in Figures \ref{fig:examplePairs} and \ref{fig:sourroundingExample}. Yet, when looked at in isolation, the clicks and echoes are too variable, as can be seen in Figure \ref{fig:exampleclicks}. 
Features like peak and average energy, phase, and structure alone are insufficient for differentiation between clicks and echoes. 
To achieve accurate differentiation between clicks and echoes, these characteristics should be analyzed in conjunction with other metrics, such as context information like inter-arrival times, as well as the differences in peak values, averages, and the structural patterns of multiple events.

\subsubsection{Other Noises} 
Unfortunately, pulsed Dirac-like impulses, such as clicks, are common in nature. 
After all, the killer whale is not the only whale that uses echolocation. 
In addition, other sources such as raindrops, boat noises, or technical interferences like mechanical or electrical noise can also produce different Dirac-like impulses which aren't always pulsed, but are still similar enough to be difficult to discern from clicks as can be seen in Figure \ref{fig:Noises}. 
\begin{figure}[!htb] \centering
\begin{subfigure}[t]{0.30\textwidth}
  \includegraphics[width=\linewidth]{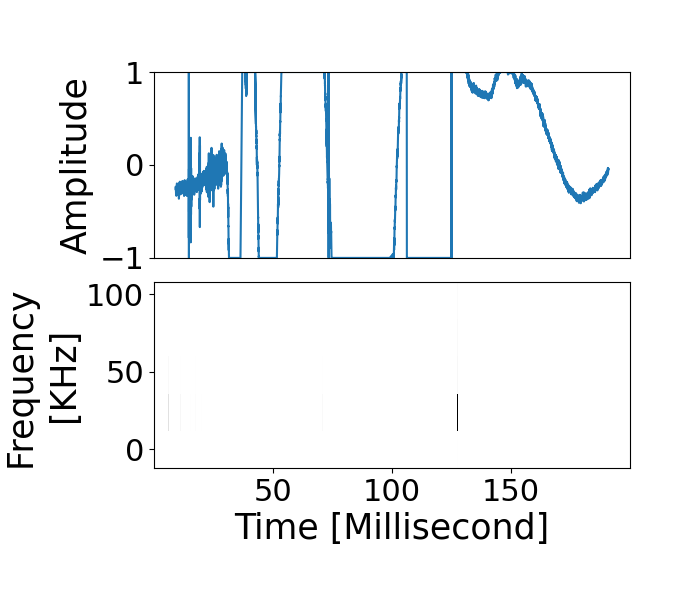}
  \caption{Waveform and spectrogram of an interference}
\end{subfigure}\hfil 
\begin{subfigure}[t]{0.30\textwidth}
  \includegraphics[width=\linewidth]{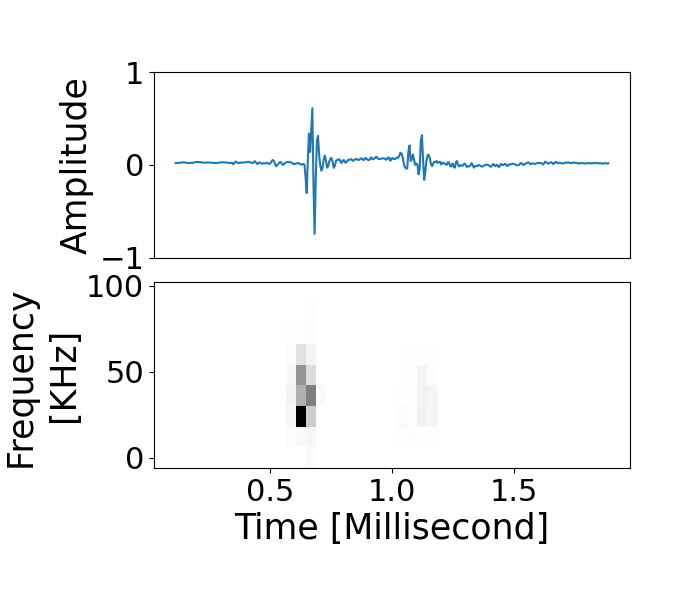}
  \caption{Waveform and spectrogram of a dolphin click}
\end{subfigure}\hfil 
\begin{subfigure}[t]{0.30\textwidth}
  \includegraphics[width=\linewidth]{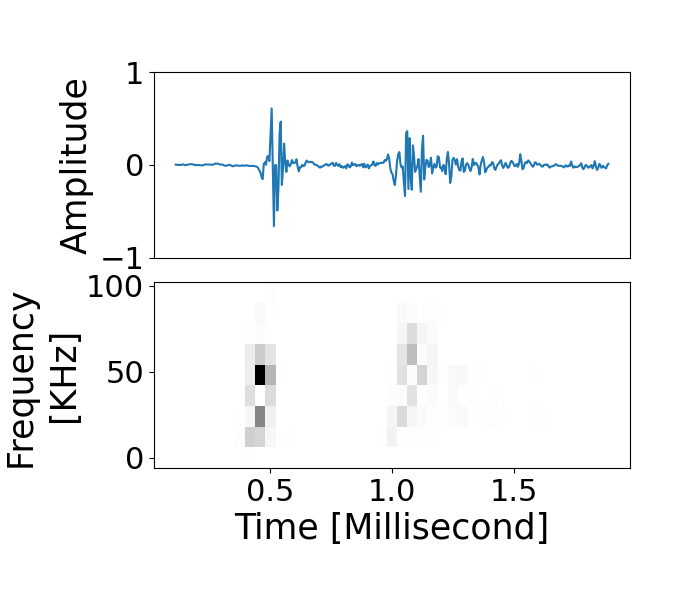}
  \caption{Waveform and spectrogram of a pilot whale click}
\end{subfigure}
\caption{Three examples of possible noises. Image (a) shows mechanical or electrical interference. Image (b) depicts a click and echo pair from a dolphin, image (c) from a pilot whale.}
\label{fig:Noises}
\end{figure}
These interferences can be a challenge, when distinguishing those from killer whale clicks, especially when the events are reverberated and attenuated. Consequently, simple mathematical assumptions are insufficient for reliable differentiation between clicks, echoes and noise in such complex underwater environments.

\subsection{First Order Gradient Conversion}
As clicks and echoes are Dirac-like in nature, they usually possess a steeper gradient in comparison to their surroundings. 
As such, the gradient of an audio signal could be used as a good feature for click and echo event detection. 
The first order gradient conversion can be used to transform the signal into its gradient representation by subtracting a value from its previous value. It is identical to the first step of the continuous wavelet transformation of the Haar wavelet. 
\begin{figure}[!htb] \centering
\begin{subfigure}[t]{0.45\textwidth}
  \includegraphics[width=\linewidth]{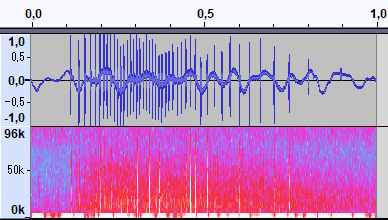}
  \caption{The original signal represented in its waveform and spectrogram. }
\end{subfigure}\hfil 
\begin{subfigure}[t]{0.45\textwidth}
  \includegraphics[width=\linewidth]{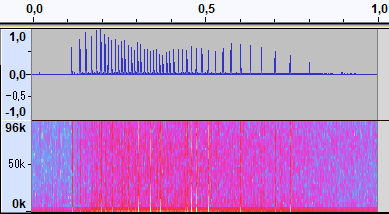}
  \caption{The first order gradient converted signal. The spectrogram shows that a lot of noise is still present in between the gradient peaks. }
\end{subfigure}\hfil 
\begin{subfigure}[t]{0.45\textwidth}
  \includegraphics[width=\linewidth]{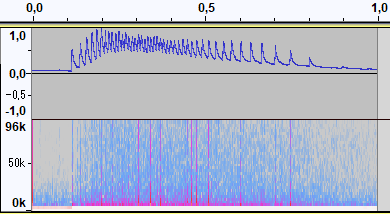}
  \caption{Moving average. The local gradient average over 1000 samples for every sample.  }
\end{subfigure}\hfil
\begin{subfigure}[t]{0.45\textwidth}
  \includegraphics[width=\linewidth]{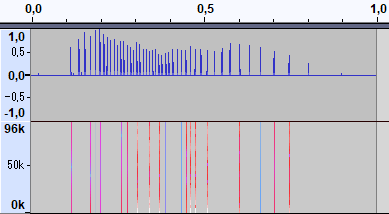}
  \caption{Noise reduction. The moving average and samples are calculated against a threshold function.}
\end{subfigure}\hfil 
\caption{This figure was generated using Audacity. 
It displays the process of the first order Dirac-like impulse detection (FOD). Image (a) shows an example of an original signal. Image (b) displays the absolute values of the first order of the signal. Image (c) is a moving average over the signal, which is used to increase or reduce a threshold function. Image (d) shows the noise-reduced Dirac-like remains of the absolute values of the first order signal. Some of the remains are so short that they do not appear on the spectrogram. }
\label{fig:FOD_Progression}
\end{figure}
The first-order gradient conversion does not provide a clean transformation of the signal. To effectively extract the gradient peaks, it is necessary to remove the noise. However, due to factors such as distance, reverberation, and attenuation, a simple threshold is insufficient for effective noise suppression. Instead, the local mean energy fluctuations of the signal must be considered when determining the noise removal threshold. To account for this, a moving average function over 1000 local gradients was applied. \\
The noise reduction equation:
\begin{equation}
Nr(s) = \left\{ \begin{array}{rcl}
s & \mbox{for} & s\geq 8m^2+2.4m+0.024 \\
0 & \mbox{for} & s < 8m^2+2.4m+0.024 
\end{array}\right.
\label{eq:Noise}
\end{equation}
Nr(s) represents the noise-reduced gradient sample, where s is the gradient sample of the converted signal, and m is the local moving average of the sample (calculated over 1000 samples). This equation was derived by fitting a curve to a test dataset, and it applies to the first-order noise-reduced Dirac-like impulse detector (FOD).
Using the noise reduction equation, the moving average and the gradient of the signal are calculated against a threshold function to transform the gradient signal into a noise-reduced Dirac-like impulse peak representation. 
This process can be seen in Figure \ref{fig:FOD_Progression}.
The first order gradient can also be padded to gain the gradient for lower frequencies. 
In this work, only the unpadded first order gradient conversion is used. This application of the FOD can also be used as a standalone detector. 

\subsection{Spectrogram and Phase}
The spectrogram \cite{Müller15} is one of the most widely used visual representations of audio signals. 
It is derived from the Short-Time Fourier Transform (STFT), which analyzes the signal in small, overlapping time segments.

\subsubsection{From the Short Time Fourier Transformation to the Spectrogram}
The STFT performs spectral analysis on a time window of the target signal \cite{Müller15}. 
The window size is typically a power of 2, which corresponds to the number of frequency bins, representing the number of overlapping sinusoidal waves used to approximate the signal inside the time window.
The STFT calculates two components for each frequency bin, the in-phase component and the out-of-phase component. 
These are returned as a complex number, with the in-phase component corresponding to the real part and the out-of-phase component corresponding to the imaginary part. 
By combining both components, we can calculate the magnitude, which represents the strength (as in amplitude) of the signal as the absolute value of the complex number, and the phase, which corresponds to the angle between the real and imaginary components. 
The magnitude is represented as a one-dimensional array of amplitudes across the frequency bins, reflecting the frequency spectrum of the signal. 
This array can then be transformed into a 1D pixel array, representing the time window.
This process of generating 1D pixel arrays is repeated multiple times by shifting the window with a hop over the signal. 
For each window, a new 1D pixel array is generated, which are combined into a 2D image, which is known as the spectrogram.

\begin{figure}[!htb] \centering
\begin{subfigure}[t]{0.25\textwidth}
  \includegraphics[width=\linewidth]{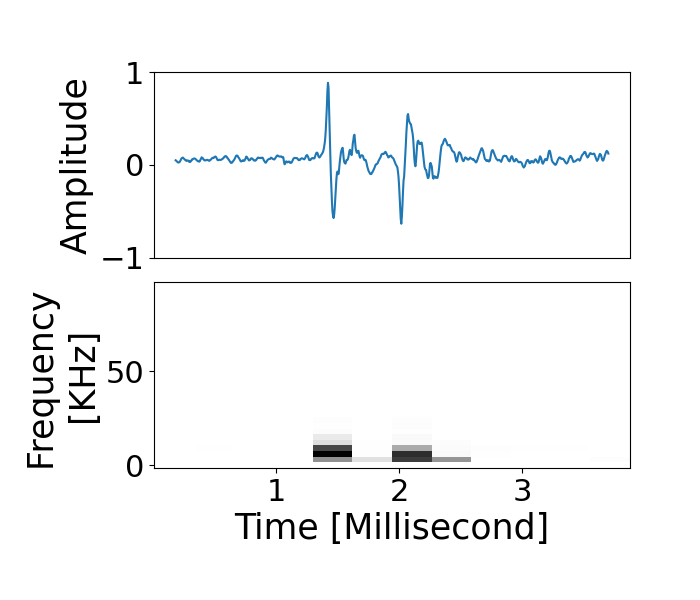}
  \caption{Waveform and spectrogram with window size 64 samples \\(0.3 Milliseconds).}
\end{subfigure}\hfil 
\begin{subfigure}[t]{0.25\textwidth}
  \includegraphics[width=\linewidth]{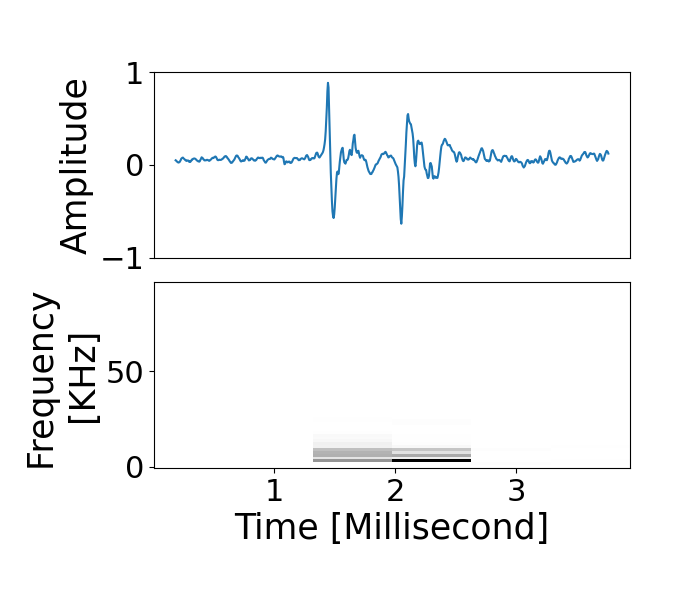}
  \caption{Waveform and spectrogram with window size 128 samples \\(0.7 Milliseconds).}
\end{subfigure}\hfil 
\begin{subfigure}[t]{0.25\textwidth}
  \includegraphics[width=\linewidth]{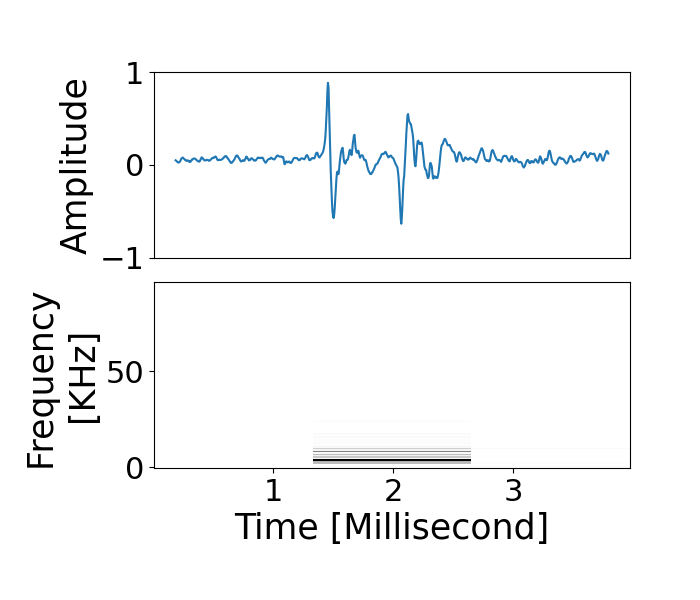}
  \caption{Waveform and spectrogram with window size 256 samples \\(1.3 Milliseconds).}
\end{subfigure}
\caption{Three examples of the same click and echo with different spectrogram window sizes. Image (a) shows the waveform and spectrogram of a signal with an FFT window size of 64. With this window size, the click and echo can still be distinguished, although there is a poor tradeoff between frequency and time resolution. Image (b) uses a window size of 128 samples. With this larger window, it becomes difficult to differentiate between the click and the echo, and pinpointing their exact start or end is challenging. Finally, Image (c) uses a window size of 256 samples, where the click and echo have merged due to the reduced time resolution, thus making them indistinguishable.}
\label{fig:uncertainty}
\end{figure}

\paragraph{The Uncertainty Principle}
The uncertainty principle concerning the spectrogram states that either a high-frequency resolution or a high time resolution can be achieved, depending on the window size, but not both simultaneously.
A larger window provides more frequency bins, resulting in higher frequency resolution. 
However, the larger window also blurs the time resolution.
Conversely, a smaller window size offers a better time resolution, but because it allows for fewer frequency bins, the frequency resolution decreases.
This problem is depicted in Figure \ref{fig:uncertainty}.
It is possible to interpolate values in both the time and frequency domain to generate additional pixels, but this interpolation does not improve the resolution. 
Another option for generating more pixels in the time domain is to reduce the hop size. 
However, a larger window will still blur the time resolution over multiple pixels, even with a smaller hop size.
The choice of window size is typically a trade-off between frequency and time resolution, depending on the specific task at hand. 
For example, in the case of a Dirac-like click sound, which is a brief signal that spans the entire frequency range, time resolution becomes more important than frequency resolution. 
If time resolution is blurred, it could be difficult to accurately pinpoint the start and end of the signal.

\subsubsection{The Phase of a Spectrogram}
The phase is the angle between the real and the complex part of the spectrogram. It provides further information on the angle of the waveform, which, in addition to the amplitudes of the spectrogram, can be used so that the unique wave coefficients can be reconstructed.

\subsection{Wavelet and Scalogram}
The wavelet transformation \cite{Bolós2014,PyWavelets} is similar to filter transformations, such as Gaussian or Laplace filters. Even though multidimensional wavelets exist, this work focuses solely on one-dimensional audio signal streams to generate a 2D image representation also known as a scalogram. Therefore, in this thesis only one-dimensional wavelet transformations are discussed.

\subsubsection{The Wavelet}
\begin{figure}[!htb] \centering
\begin{subfigure}[t]{0.45\textwidth}
  \includegraphics[width=\linewidth]{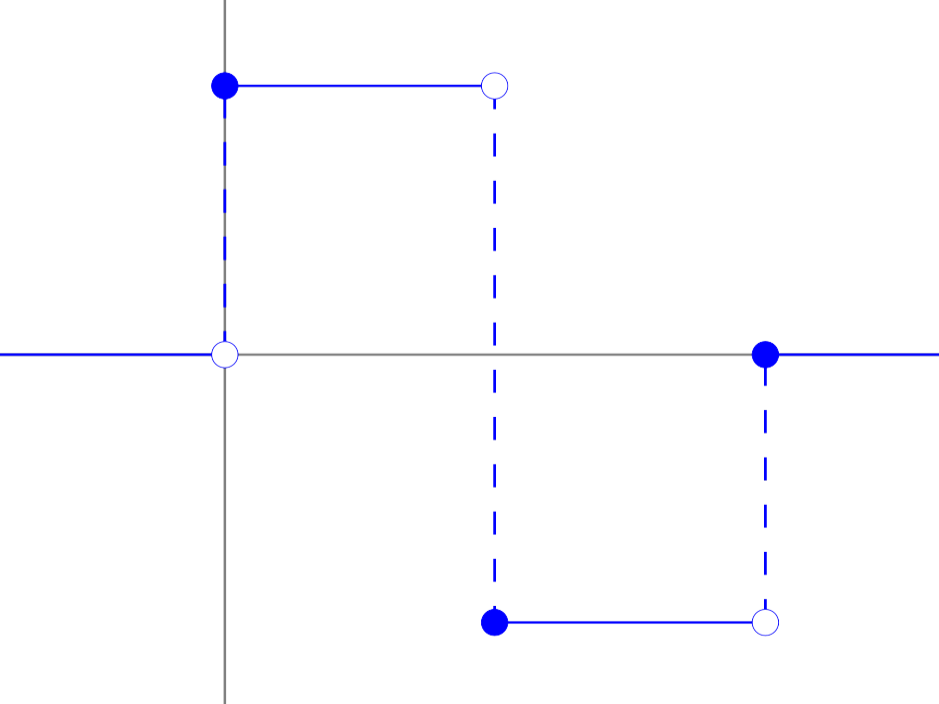}
  \caption{The discontinuous, unsymmetrical Haar wavelet. }
\end{subfigure}\hfil 
\begin{subfigure}[t]{0.45\textwidth}
  \includegraphics[width=\linewidth]{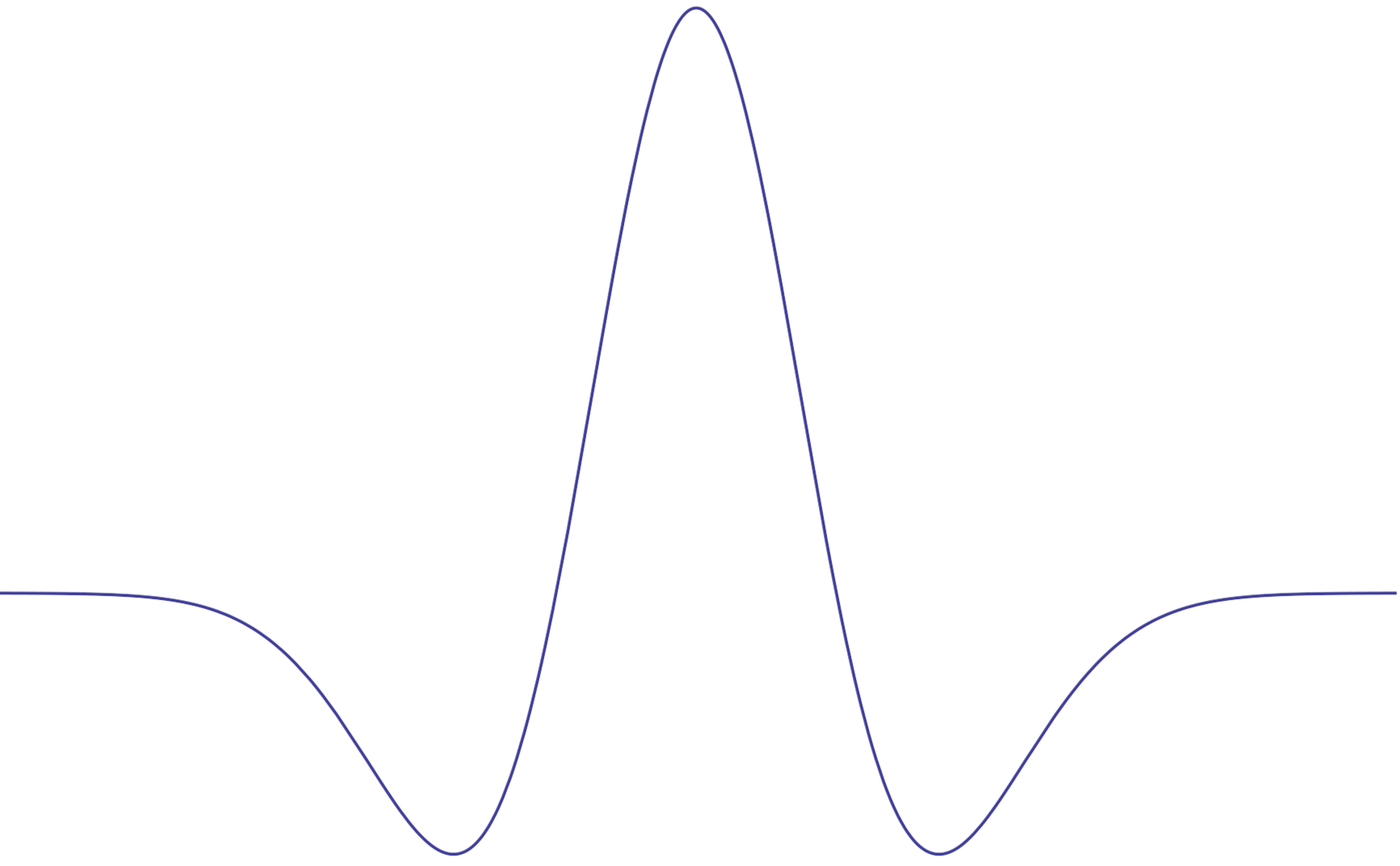}
  \caption{The continuous, symmetrical Mexican Hat wavelet.}
\end{subfigure}\hfil 
\caption{Examples of wavelets taken from Wikipedia \cite{Wavelet} The first step of the Haar continuous wavelet transformation of image (a) was used for the first order gradient conversion and first order Dirac-like event detection (FOD). The Mexican Hat wavelet from image (b) was used to generate the scalogram. The Mexican Hat is the smallest symmetrical wavelet in terms of used samples.}
\label{fig:Examplewavelets}
\end{figure}

Unlike the infinite sinusoidal filters used in the STFT transformation, wavelet filters are finite in length. A wavelet can be thought of as a window function that transforms a signal into a filtered representation using the Continuous Wavelet Transform (CWT) function. Typically, the window is shifted by one sample at a time, which ensures that the filtered representation has the same sample size as the original signal. By increasing the window size, the resulting representation becomes smaller. There are several types of wavelets, such as the Haar wavelet, Mexican Hat wavelet, Meyer wavelet, and Morlet wavelet, to name a few. The Haar wavelet and Mexican Hat wavelet are both depicted in Figure \ref{fig:Examplewavelets}. In this work, after careful consideration, the Mexican Hat wavelet was chosen for the CWT function to transform the signal. This decision was based on the fact that the Mexican Hat is a compact and symmetrical wavelet which performs the second order conversion, making it particularly effective for detecting higher frequency changes.

\paragraph{Scaling}
A single wavelet transformation is not enough to transform the signal into a 2D representation. 
The window size of a wavelet corresponds to the frequency range it represents. 
By increasing the window size, the frequency range covered by the transformation changes. 
The window size of a wavelet can be adjusted through scaling.
Each scaled version of the wavelet represents a different frequency range, and each version corresponds to a pixel on the scalogram. 
Scaling the wavelet also affects the time-frequency resolution. A smaller wavelet window provides low-frequency resolution but high time resolution for high frequencies, while a larger window offers a high frequency resolution for low frequencies but lower time resolution.

\subsubsection{Scalogram}
\begin{figure}[!htb] \centering
\includegraphics[width=0.9\linewidth]{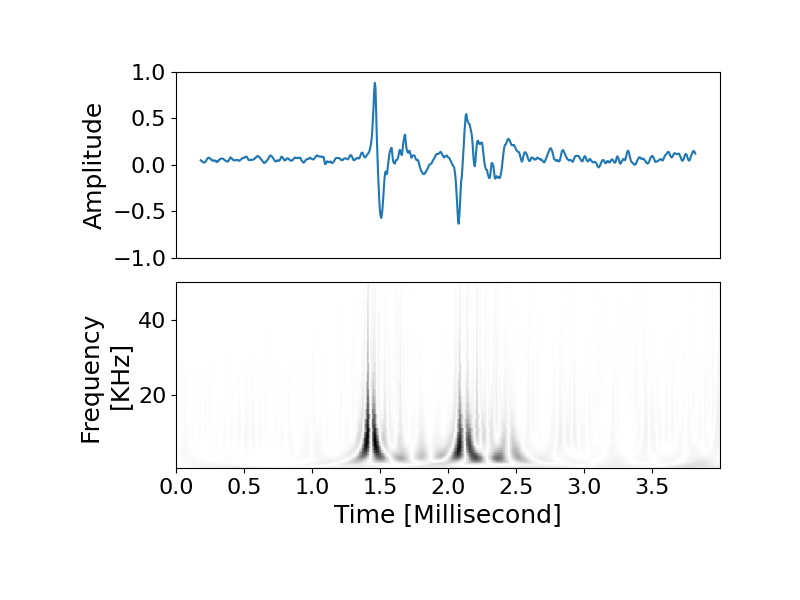} 
\caption{An example of a Scalogram with its respective waveform. The Dirac-like impulses can be easily seen and tracked as the outlying cones protruding from the top of the image. The representation can be particularly useful for a click detector. }
\label{fig:exampleScalogram}
\end{figure}

By stacking the scaled wavelet transformations, a 2D image representation of the signal is generated based on the scaled wavelets. An example of a scalogram is depicted in Figure \ref{fig:exampleScalogram}.  
The frequency axis follows a logarithmic distribution, where higher frequencies have high time resolution and lower frequencies have high-frequency resolution.
The main advantage of the scalogram, compared to the spectrogram, is that Dirac-like impulses are better visualized at high frequencies, and the scalogram also provides a denoising effect. 
However, similar to the spectrogram, the phase information is lost in the wavelet representation.
Overall, the denoising effect and high-frequency time resolution make the scalogram particularly useful for a deep learning-based click detector. In order to improve performance, the scalogram was built using 20 values from a geom space between 1 and 50. This means that the scalogram represents a frequency range between 960 Hz and 48 kHz. This means that frequencies higher than 48 kHz are not depicted. Yet, the gradient energy of higher frequencies are still present in the scalogram in lower frequencies. They look like protruding double cones that appear from the top of the scalogram, as can be seen in Figure \ref{fig:exampleScalogram}.

\subsection{Machine Learning}
With the advent of more powerful computers, increasingly complex tasks can be tackled. However, these tasks often cannot be fully defined through mathematical assumptions alone. This is where supervised machine learning comes in, allowing algorithms to improve by learning from prior experiences. In supervised learning, the training data is pre-labeled with a ground truth label, enabling a direct comparison between the model's predictions and the annotated results.
This section provides an overview of the key components of deep neural networks and supervised machine learning, laying the foundation for a clearer understanding of the networks and models presented in the subsequent methodology chapter.

\subsubsection{Perceptron}
\begin{figure}[!htb] \centering
\includegraphics[width=0.7\linewidth]{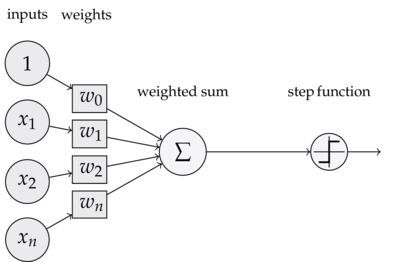}
\caption{A visual representation of the Rosenblatt perceptron using TikZ \cite{PerceptronTikZ}. The inputs X are multiplied by their corresponding trainable weights W, and the weighted sum is calculated. This sum is then passed through an activation function to produce the final output.}
\label{fig:Perceptron}
\end{figure}

\begin{figure}[!htb] \centering
\begin{subfigure}[t]{0.3\textwidth}
  \includegraphics[width=\linewidth]{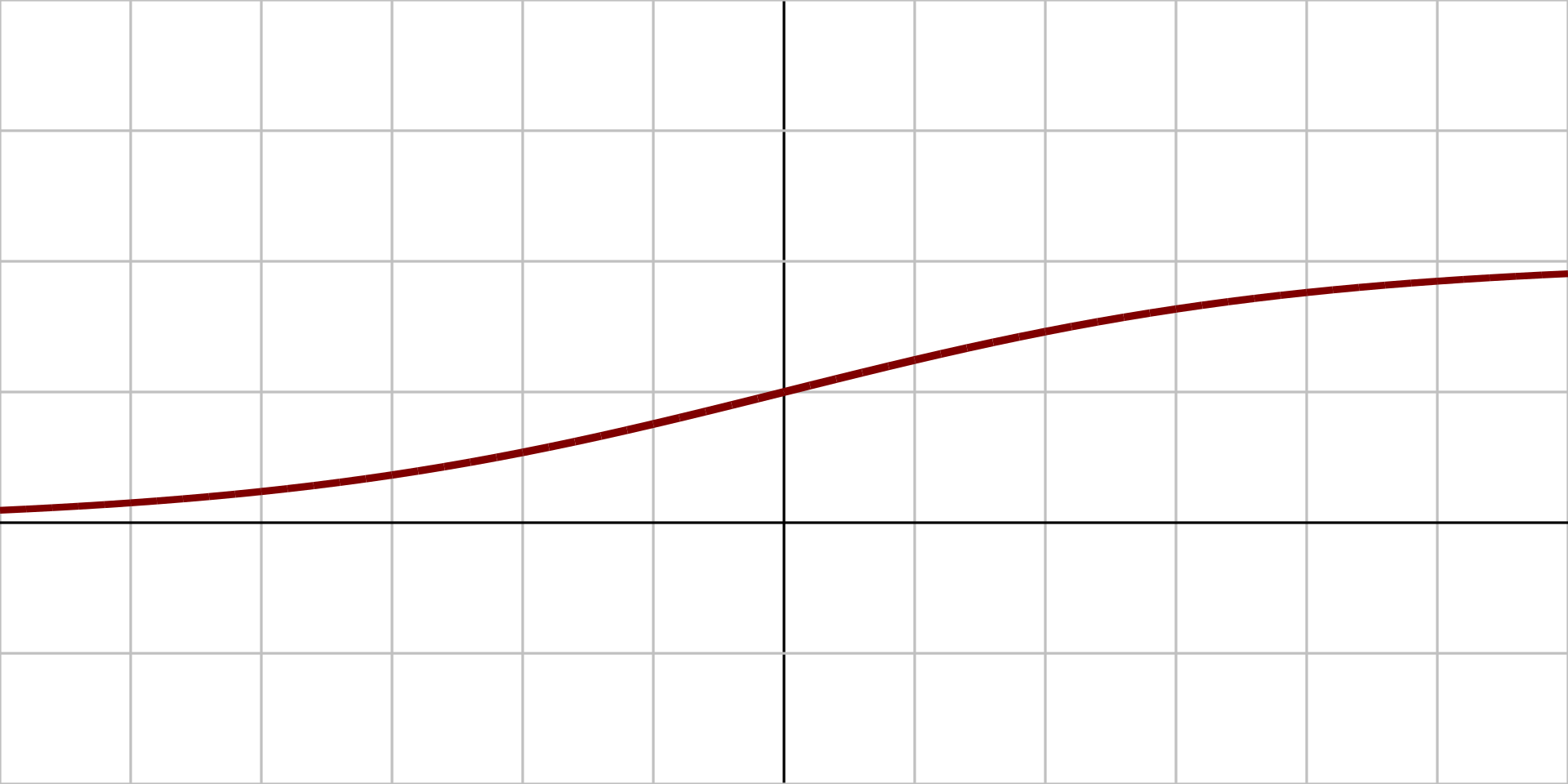}
  \caption{Sigmoid activation function}
\end{subfigure}\hfil 
\begin{subfigure}[t]{0.3\textwidth}
  \includegraphics[width=\linewidth]{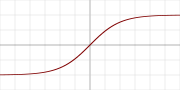}
  \caption{Tanh\\ activation function}
\end{subfigure}\hfil 
\begin{subfigure}[t]{0.3\textwidth}
  \includegraphics[width=\linewidth]{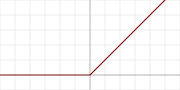}
  \caption{ReLU\\ activation function}
\end{subfigure}
\caption{Depiction of the Sigmoid(a), Tanh(b) and ReLU(c) activation functions taken from Wikipedia.}
\label{fig:activationFunctions}
\end{figure}

The perceptron \cite{Rosenblatt:1958} is a fundamental building block of neural networks and consists of four key components: the input, trainable weights, a non-linear activation function, and the output.
In operation, each input X is multiplied by its corresponding trainable weight W. The sum of these weighted inputs is then passed through an activation function to compute the perceptron's output. A depiction of the Rosenblatt perceptron can be seen in Figure \ref{fig:Perceptron}.
The activation function must be non-linear. Otherwise, a stack of perceptrons could be reduced to a single matrix-matrix multiplication, which would limit the model's ability to solve more complex tasks.
Common examples of non-linear activation functions include the binary step, sigmoid, tanh, and the Rectified Linear Unit (ReLU), along with variants such as leaky ReLU, Exponential Linear Units (ELU), and softmax, among others. Depictions of the Sigmoid, Tanh and ReLU can be seen in Figure \ref{fig:activationFunctions}.

\subsubsection{Fully Connected Neural Network Model}
\begin{figure}[!htb] \centering
\includegraphics[width=0.9\linewidth]{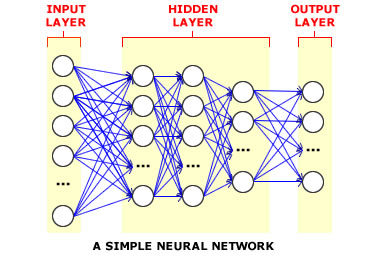}
\caption{A depiction of a small fully connected deep neural network taken from stackexchange \cite{SimpleDeepNeuralNetwork}, where every output from a previous layer is connected to the input of each perceptron in the following layer.}
\label{fig:LNN}
\end{figure}

A neural network is a model made up of stacked perceptron layers. 
In a fully connected neural network, every output from a previous layer is connected to the input of each perceptron in the following layer. A model depiction of a fully connected neural network can be seen in Figure \ref{fig:LNN}.
The network learns by adjusting the perceptron weights during training to minimize the error between the predicted and the annotated outputs over multiple examples.

\subsubsection{2D Convolutional Neural Network}
\begin{figure}[!htb] \centering
\includegraphics[width=0.9\linewidth]{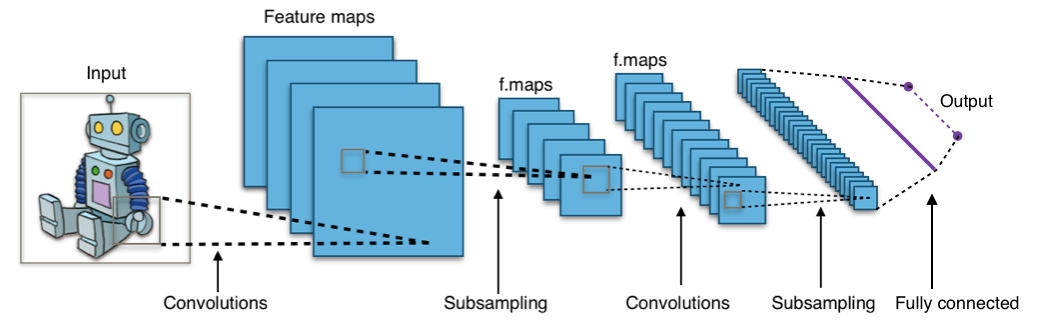} 
\caption{Depiction of a typical deep CNN model. In the first layer, the original image is transformed into multiple representative feature maps, every feature map is a convolution of the original image over a kernel filter. This step is repeated iteratively, the subsequent 2D convolutions in deeper layers are stacked on the feature maps generated by earlier layers until the feature map is flattened into a 1D array and further processed using a fully connected neural network. The image was taken from Park explanation of CNN models\cite{Park20}.}
\label{fig:CNN}
\end{figure}

A fully connected neural network performs optimally when the inputs are independent of each other. However, this is not always the case. In images, pixels that are close together are more likely to be related to a common shape than those that are farther apart. In such scenarios, convolutional neural networks (CNNs) are designed to exploit the spatial locality of data to extract meaningful features. While convolutions can exist in higher dimensions, this thesis focuses on 2D convolutions for image processing.
\begin{figure}[!htb] \centering
\includegraphics[width=0.9\linewidth]{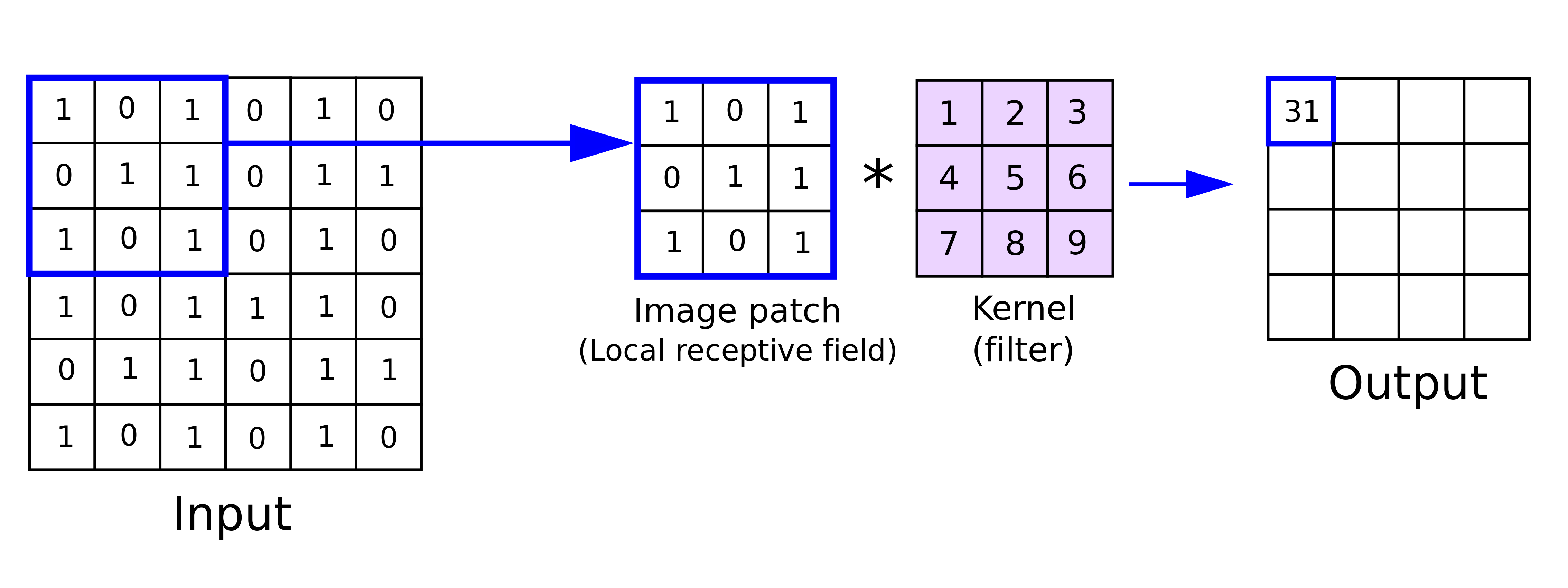} 
\caption{Depiction of a 2D convolution with a stride of 1 and no padding. The original 6x6 image is processed into a smaller 4x4 representation using a 3x3 kernel. The image was taken from zaforf GitHub \cite{zaforf}.}
\label{fig:2Dconv}
\end{figure}
A 2D convolution is a type of perceptron where the inputs are arranged into a filter, typically a square filter such as 3x3, 5x5, or 7x7. These trainable filters help generate a new representation of the original image based on the perceptron’s learned weights. Figure \ref{fig:2Dconv} depicts how a 2D convolution transforms an image into a feature map representation.
Additionally, these convolutional filters can be stacked, allowing subsequent 2D convolutions in deeper layers to operate on the feature maps generated by earlier layers, as depicted in Figure \ref{fig:CNN}.
Moreover, convolution can be used to reduce the spatial dimensions by increasing the hop or stride of the image while preserving the essential features and relevant information extracted by the filters.

\subsubsection{Network Training}
Supervised training of a neural network typically requires three datasets comprised of annotated data.
A training dataset, a validation dataset and a test dataset.
The training dataset is used to iteratively improve the network in two steps. 
First, the training data is fed into the network to calculate an output. This output is then compared to the expected output of the training data, known as the label, to calculate the loss of the network on the sample. 
In the second step, a small learning rate is used to backpropagate the loss through the network. This process adjusts the trainable weights, bringing the output closer to the expected value. These steps are repeated over multiple iterations of the entire training set, also known as an epoch, to refine the network’s performance.
The validation dataset is used to monitor the network’s progress and ensure it is improving. Sometimes, instead of learning the task at hand, the network may start memorizing specific patterns or characteristics of the training data that are not related to the task, a problem known as overfitting. To mitigate overfitting, the validation dataset is passed through the network after each iteration to calculate the loss, but without backpropagation. If the loss on the validation dataset stops decreasing, it indicates that the network has stopped improving on the task and may be overfitting on the training data.
Finally, the test dataset is used after training to evaluate the network's performance on unseen data, providing insight into its generalization ability and overall effectiveness.

\subsection{Decision Trees}
Due to the similarities between clicks and echoes, it can be challenging to distinguish between them based on isolated events alone. Biologists, however, use the surrounding context of an event to aid in labeling. By comparing an event with its neighboring context, it becomes easier to identify whether the event is a click or an echo. To replicate this approach, the results of isolated event detection can be enriched with contextual information and then processed through a decision tree.
\begin{figure}[!htb] \centering
\includegraphics[width=0.9\linewidth]{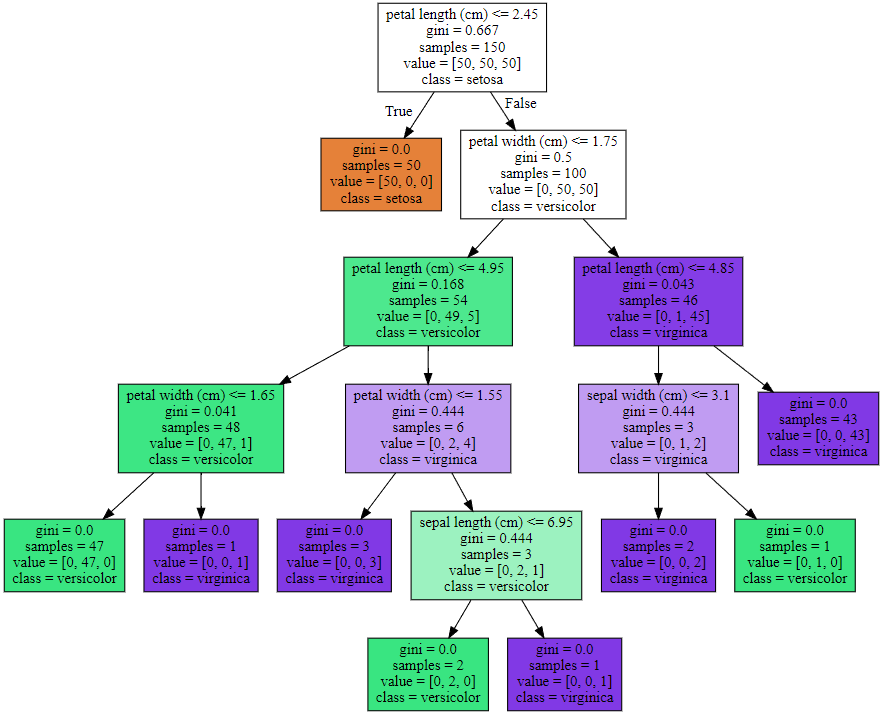}
\caption{A visualization of an example decision tree based on the iris dataset using sklearn and graphviz. This visualization was taken from the scikit-learn examples \cite{scikit-learn}.The Iris dataset is a well-known introductory dataset used for classifying three species of flowers (Setosa, Versicolor, and Virginica) based on the measurements of their petal and sepal width and length. }
\label{fig:DecisionTree}
\end{figure}
A classification decision tree is a supervised learning method in which the leaves of the tree represent class labels, while the branches correspond to combinations of features that lead to those labels \cite{Quinlan86}. Figure \ref{fig:DecisionTree} visualizes an example decision tree. 
Decision trees are trained by selecting the best feature to split the dataset into subsets, using criteria such as information gain and Gini impurity. Information gain measures the reduction in uncertainty after a split, while Gini impurity quantifies the distribution of samples from different classes within a node. A pure node, where all samples belong to a single class, is considered a leaf in the decision tree and has a Gini impurity of zero. During tree construction, the feature that maximizes information gain and minimizes Gini impurity is chosen for the split. This process is repeated recursively until one of the stopping conditions is met: the tree reaches a predefined depth, a node contains fewer than a minimum number of samples, or no further significant reduction in impurity is possible.

\section{Methodology}
The following section gives a detailed overview of the corresponding methodologies employed within this work, such as the tools utilized during the experimentation, as well as the settings and changes applied. 
No animals were directly involved in this study.

\subsection{Sound Segmentation using Deep Learning with ANIMAL-SPOT}
\begin{figure}[!htb] \centering
\includegraphics[width=\linewidth]{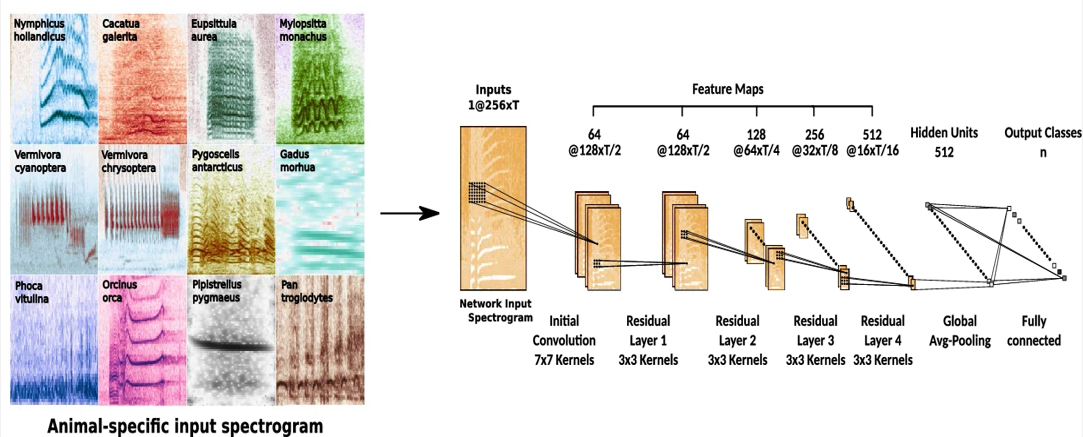}
\caption{Depiction of the ANIMAL-SPOT model architecture and spectrograms of some of the species' target calls taken from the ANIMAL-SPOT publication by Bergler et al. \cite{Bergler21-AAA}}
\label{fig:ANIMAL-SPOT}
\end{figure}

\begin{figure}[!htb] \centering 
\begin{subfigure}[t]{0.33\textwidth}
  \includegraphics[width=\linewidth]{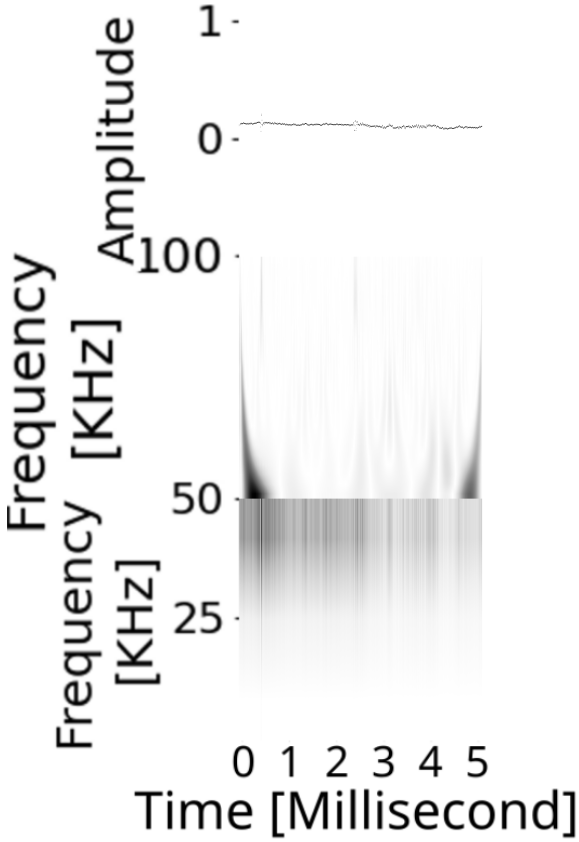}
  \caption{}
\end{subfigure}\hfil 
\begin{subfigure}[t]{0.33\textwidth}
  \includegraphics[width=\linewidth]{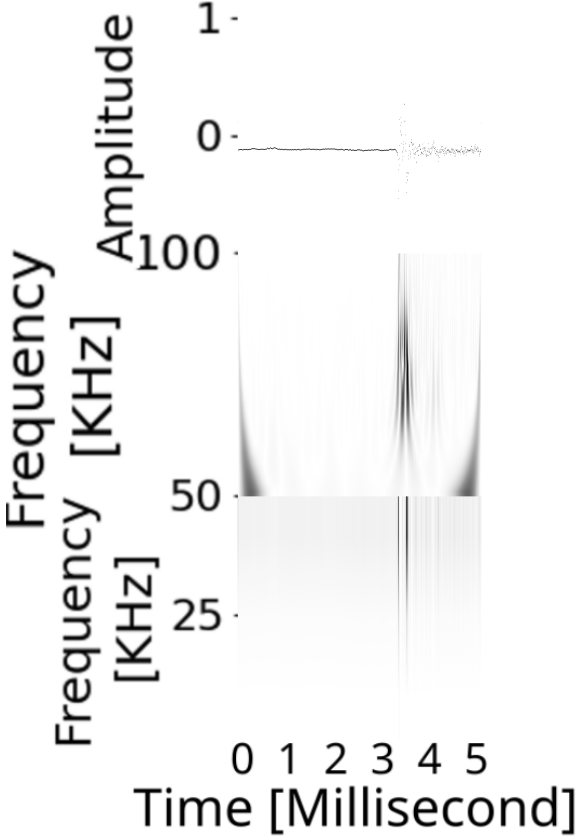}
  \caption{}
\end{subfigure}\hfil 
\begin{subfigure}[t]{0.33\textwidth}
  \includegraphics[width=\linewidth]{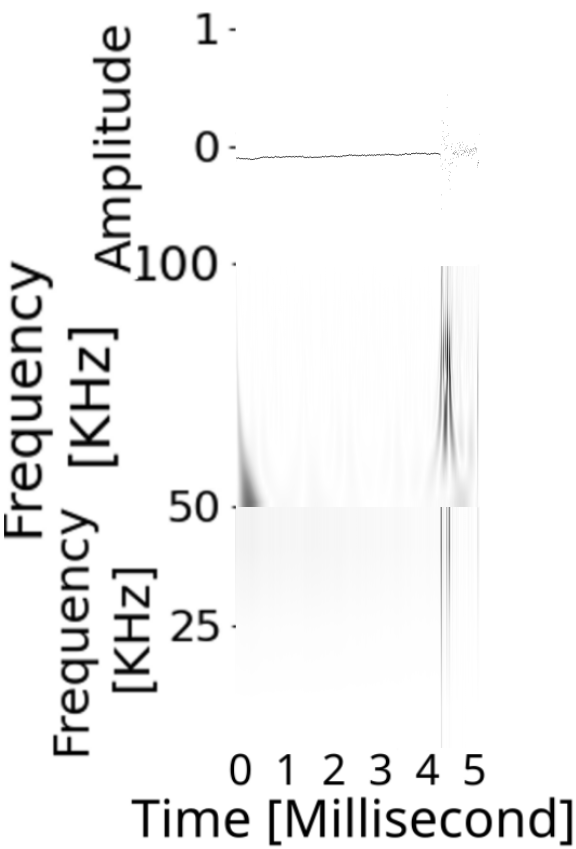}
  \caption{}
\end{subfigure}
\caption{ Three example depictions of the adjusted ANIMAL-SPOT input. To simplify the viewing, the waveform (top), CWT (middle) and spectrogram (bottom) images were stacked vertically and not over the color channels. A channel stacked example can be found in Figure \ref{fig:SCWTSPEC}.}
\label{fig:ANIMAL-SPOT-Adjusted-Input}
\end{figure}

ANIMAL-SPOT \cite{Bergler21-AAA} is a windowed deep learning classifier based on the ResNet-18 architecture. It is capable of both binary event detection and multi-class classification. 
The model has been successfully trained on audio data from various species for different research projects, including calls from cockatiels, cockatoos, conures, monk parakeets, warblers, penguins, Atlantic cod, harbor seals, killer whales, pygmy pipistrelles, and chimpanzees \cite{Bergler21-AAA}.
A depiction of the ANIMAL-SPOT architecture and spectrograms of the target input signals can be seen in Figure \ref{fig:ANIMAL-SPOT}.
The model transforms a window of input audio data into spectrograms and then performs image recognition on these spectrograms. 
During prediction, the trained ANIMAL-SPOT model processes the audio file by splitting the data stream into fixed-size windows. 
The window size is selected to be large enough to capture the majority of the target signals, but small enough that, on average, only one signal fits within each window. 
The window is typically shifted by a hop size equal to half the window length, resulting in a 50\% overlap between adjacent windows. 
ANIMAL-SPOT calculates a certainty score (ranging from 0 to 1) for each window, indicating the likelihood of a specific class being present in the window. 
This certainty score is compared to a predefined threshold, if no class exceeds the threshold, the window is labeled as noise. 
If one or more classes exceed the threshold, the class with the highest certainty is assigned to the window, or all classes above the threshold are assigned, dependent on the task.
For this thesis, ANIMAL-SPOT's preprocessing was adapted to include waveform and continuous wavelet transform (CWT) image representations, in addition to the standard spectrogram. 
This modification allows the preprocessing pipeline to generate signal, CWT and spectrogram images, as depicted in Figure \ref{fig:ANIMAL-SPOT-Adjusted-Input} from the audio data.

\subsection{Object Detection for Acoustic Event Recognition with YOLO}
YOLO (You Only Look Once) \cite{Redmon15} is a convolutional neural network with fast training and inference time \cite{Redmon18}. 
The YOLO network performs both the detection of bounding boxes and the classification by dividing the input image into a grid of cells. 
\begin{figure}[!htb] \centering
\includegraphics[width=0.9\linewidth]{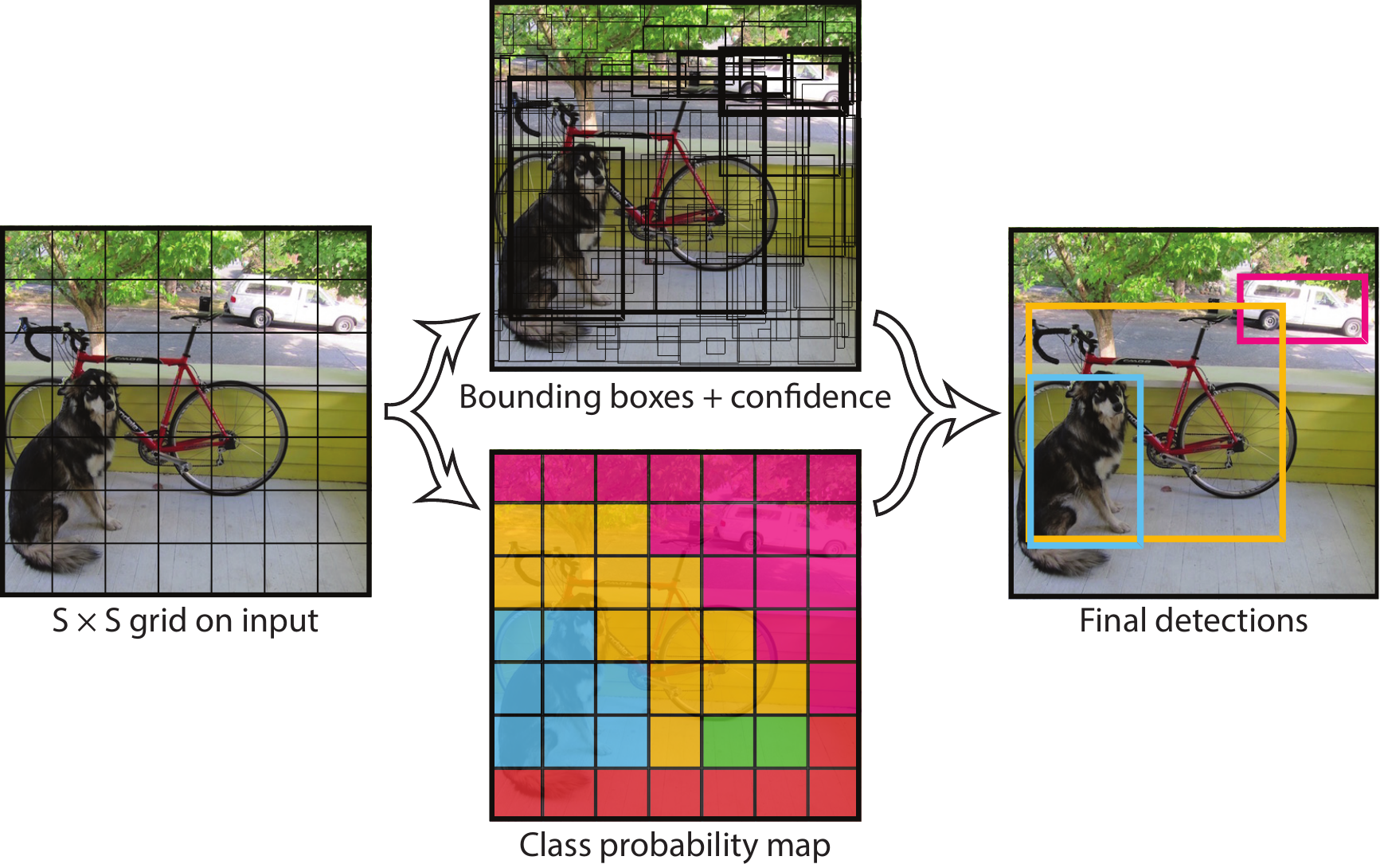}
\caption{An example of how YOLO predicts bounding boxes based on a grid of cells. The image is taken from the YOLO publication by Redmon et al. \cite{Redmon15}. The model divides the image into a grid and for each
grid cell it predicts bounding boxes, confidence for those boxes,
and class probabilities.}
\label{fig:Yolo_Example}
\end{figure}
Each cell is responsible for detecting objects within its region by predicting bounding boxes and class probabilities. 
These predictions are then combined using non maxima suppression (NMS) based on the intersection-over-union (IoU) metric to calculate the final set of bounding boxes per object detection. 
Object classification is determined by a confidence threshold applied to these bounding boxes. 
A simplified workflow can be seen in Figure \ref{fig:Yolo_Example}.
Over time, YOLO has evolved through various versions, from the original network based on GoogleNet or VGG16 architectures to the current state-of-the-art YOLOv10 \cite{wang24}. Below is a summary of the key developments in these iterations \cite{Ultralytics}:
\begin{enumerate}
\item \textbf{YOLO} \\
The original version divided the image into a single grid (typically 7x7), due to the coarse cell size the first YOLO has issues handling larger objects that span multiple grid cells or objects that are too small.
\item \textbf{YOLOv2} \\
YOLOv2 adopts the Darknet-19 architecture \cite{YOLO9000}, which has fewer layers than the GoogleNet or VGG16 architecture. It also adds batch normalization to stabilize the learning process. In addition, YOLOv2 uses anchor boxes to simplify the bounding box generation and adds the ability to find multiple bounding boxes per cell. 
\item \textbf{YOLOv3} \\
YOLOv3 uses the deeper Darknet-53 architecture \cite{YOLOv3PWC}, which also has residual connections. In addition, the deeper network also introduced three different scales of granularity. A coarse scale of 13x13, medium scale of 26x26, and fine scale of 52x52. The different scales are used to improve object detection for different-sized objects. 
\item \textbf{YOLOv4} \\
YOLOv4 is designed with a stronger focus on hardware efficiency, the main additions are performance optimizations and a new cross-stage partial network (CSP) Darknet53 backbone which is optimized for GPU performance.
\newpage
\item \textbf{YOLOv5} \\
Unlike the prior models, which were developed by Joseph Redmon and his team, YOLOv5 and the following YOLO versions are developed by Ultralytics \cite{Ultralytics}. One of the new aspects is the introduction of model sizes (YOLOv5s, YOLOv5m, YOLOv5l, YOLOv5x) for different applications, which allows the users to balance speed and accuracy depending on the task. 
\item \textbf{YOLOv6} \\
YOLOv6 introduced the EfficientNet \cite{Ultralytics} backbone, an architecture with fewer parameters designed to run on lower-powered hardware, such as mobile devices or IoT systems. 
It also uses more advanced data augmentation techniques (e.g., CutMix, Mixup) and adversarial training to improve robustness.
\item \textbf{YOLOv7} \\
YOLOv7 improved upon YOLOv6 by offering significant architectural enhancements for both lower-powered edge devices and high-performance computing environments. Thus offering flexibility for real-time applications and cloud-based deployments.
\item \textbf{YOLOv8} \\
YOLOv8 is an optimized version of YOLOv7, the main improvement was YOLOv8 enhanced compatibility with more deployment frameworks, including TensorFlow Lite and PyTorch. YOLOv8 is used as a standard due to its enhanced Python compatibility.  
\item \textbf{YOLOv9} \\
YOLOv9 introduced major changes in the core structure. It uses a more shallow network with residual connections and dilated convolutions to improve information preservation and feature extraction in deep networks. 
\item \textbf{YOLOv10} \\
YOLOv10 is the current state-of-the-art version as of the time of this writing. YOLOv10 focuses on latency reduction, efficiency improvements, and increased generalization when compared to prior iterations. It achieves this by replacing the non-maximum suppression with dual assignment.
\end{enumerate}
In this work, the YOLOv8 network, provided by Ultralytics \cite{Ultralytics, yolov8_ultralytics}, was first tested for the CLICK-SPOT alpha toolchain development due to its enhanced Python compatibility. It serves as the backbone of the CLICK-SPOT toolchain. 
Furthermore, the YOLOv10 network was also tested for comparative purposes but was found to have no major detection improvements when compared to the YOLOv8 network for this task.


\subsubsection{YOLO Post Processing using FOD} \label{Box_Slicing_Solutions}
\begin{figure}[!htb] \centering
\begin{subfigure}[t]{0.9\textwidth}
  \includegraphics[width=\linewidth]{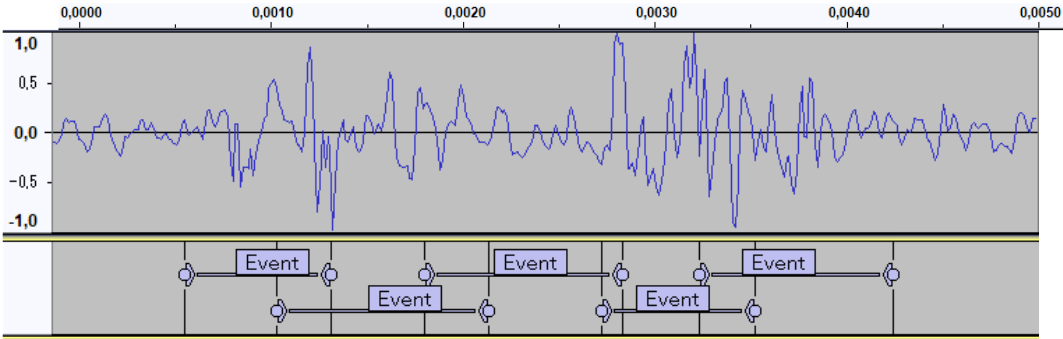}
  \caption{The raw YOLO Event Windows of a high SNR event. Due to overlap, the YOLO bounding boxes merge together into one large event, as can be seen in image (b)}
\end{subfigure}\hfil 
\begin{subfigure}[t]{0.9\textwidth}
  \includegraphics[width=\linewidth]{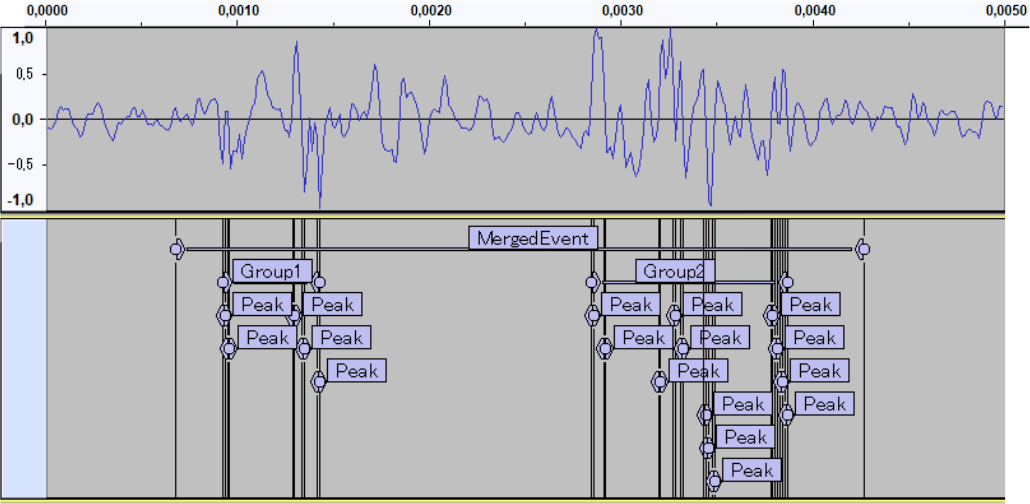}
  \caption{The FOD extracts gradient peaks from the large merged event. The peaks are grouped based on sample distance. These peak groups are forwarded to the click and echo differentiator to generate the new labels.  }
\end{subfigure}\hfil 
\begin{subfigure}[t]{0.9\textwidth}
  \includegraphics[width=\linewidth]{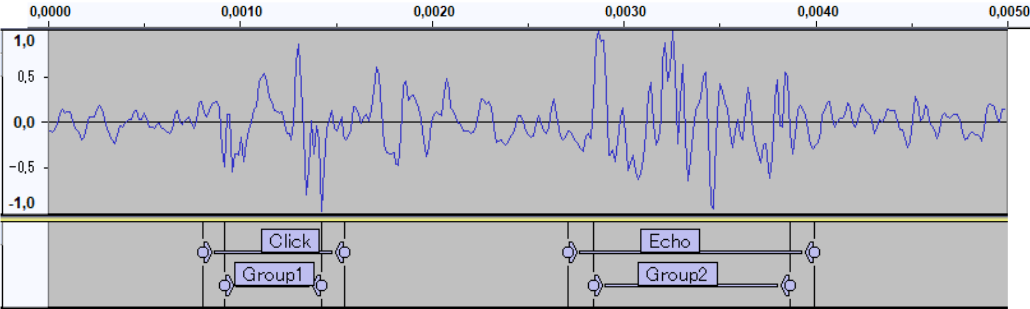}
  \caption{The resulting click and echo from the grouped gradient peaks. A padding is added to obtain more samples around the gradient peaks. The padding cannot overlap with other groups. }
\end{subfigure}
\caption{A depiction of the YOLO post-processing using the FOD. The images were made in Audacity. Figure (a) shows a scene of a low SNR signal with the corresponding YOLO bounding boxes. Due to overlap, the bounding boxes merged into a large event which includes both the click and echo. Through FOD peak detection and FOD peak grouping shown in Figure (b), the click and echo are extracted from the merged event, as can be seen in Figure (c). }
\label{fig:YOLOPostProgression}
\end{figure}

While the results from the YOLO model are more accurate than the ANIMAL-SPOT window labels, they still require refinement and conversion into usable text labels. 
Since clicks and echoes are distinctly brief and the likelihood of overlap is minimal, Dr. Vester and her team chose not to account for overlapping annotations. 
One improvement was merging bounding boxes to eliminate overlaps along the time axis, just like the provided ground truth labels. But this improvement also carried the risk of larger merged bounding boxes with both the click and echo present. To reduce this problem, the first-order gradient conversion was applied to identify the gradient peaks of the click and echo within the YOLO bounding boxes. 
These peaks can enhance detection results, as grouping them helps separate merged boxes and adjust the bounding boxes to better align with the clicks. 
A depiction of the YOLO post-processing can be seen in Figure \ref{fig:YOLOPostProgression}.

The YOLO post-procession FOD and the standalone FOD detector differ in that the local moving average of the YOLO post-procession was based on the YOLO bounding box sample size instead of 1000 samples. In all other regards, it works the same as the stand alone FOD detector. 

\subsubsection{Random Forest Post Processing}
\begin{figure}[!htb] \centering
\includegraphics[width=0.9\linewidth]{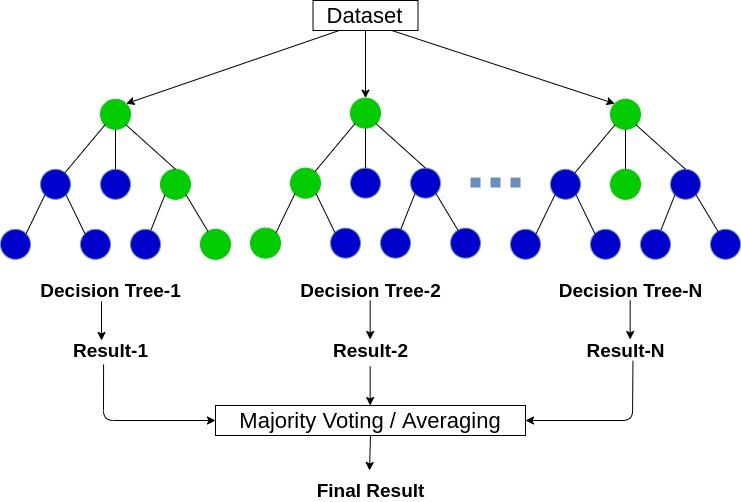}
\caption{Diagram of the random forest classifier model, showing the construction of multiple decision trees and the aggregation of their results. The example was taken from freesion \cite{RandomForestFreesion}}
\label{fig:RandomForestModel}
\end{figure}

A random forest classifier \cite{Breiman01} is a supervised ensemble learning method that constructs multiple decision trees \cite{Quinlan86}. 
The key concept behind a random forest is to train these trees using different subsets of the same training data. 
The results of all trees are then aggregated, typically by averaging their class predictions, to enhance the model’s overall performance.
Figure \ref{fig:RandomForestModel} depicts a diagram of the random forest classifier model.
In this study, the random forest classifier was utilized as a follow-up model for differentiating clicks from echoes. 
The approach leveraged the idea that by analyzing multiple results from the YOLO post-processing, along with additional contextual features—such as interarrival time, energy difference, confidence difference, and prior labels—the random forest could effectively distinguish clicks from echoes by considering events in context rather than in isolation.
While alternative models, such as a linear neural network, could have been used for this task, the random forest classifier implementation from scikit-learn \cite{scikit-learn} was chosen primarily for its simplicity in training and robustness in handling complex, high-dimensional data.

\section{Data} \label{data} 
As stated, the audio recordings used in this thesis were provided by Dr. Vester from Ocean Sounds e.V. \cite{Oceansounds}. These recordings were collected from the Norwegian west coast, specifically from the fjords near Bodø. Along with the recordings, Dr. Vester and her team also supplied hand-labeled annotations for a total of 3 minutes and 12 seconds of training material. In total, 6994 annotations were made, which are categorized into clicks and echoes. A summary of these annotations can be found in Table \ref{tab:LFHFUSECHO}.
According to time logs, the annotation process took 99 hours, 25 minutes, and 29 seconds. However, this total time includes duplications, as multiple team members annotated the same audio files to cross-check and compare their results.
\begin{table}[!ht] \centering 
    \begin{tabular}{|l||l|l||l|l|} 
    \hline
        Label & Train & Train Percentage & Evaluation & Evaluation Percentage \\ \hline\hline
        LF & 119 & 3.3 & 115 & 4.0 \\ \hline
        HF & 3491 & 96.6 & 2742 & 95.8 \\ \hline
        US & 3 & 0.1 & 6 & 0.2 \\ \hline \hline
        Clicks & 3613 & 51.7 & 2863 & 53.8 \\ \hline
        Echo & 3381 & 48.3 & 2459 & 46.2 \\ \hline\hline
        All & 6994 & ~ & 5322 & ~ \\ \hline
    \end{tabular}
    \caption{Summary of the provided labeled data. "LF" represents low-frequency clicks (below 5 kHz), "HF" refers to high-frequency clicks (between 5 kHz and 40 kHz), and "US" corresponds to ultrasonic-frequency clicks (above 40 kHz).}
    \label{tab:LFHFUSECHO}
\end{table}
The provided train data of 3 minutes and 12 seconds was divided into 38,405 input windows without overlap. Each input window has a 5 millisecond duration, equivalent to 960 samples at a sampling frequency of 192 kHz.
Every input window also has a corresponding YOLO label text file. The labels were extracted from the hand annotations and transformed into YOLO train data based on the Label, x, y, width, height encoding.
For the event detection, the clicks and echoes were not differentiated in the text label file. 

\begin{figure}[!htb] \centering
\includegraphics[width=0.6\linewidth]{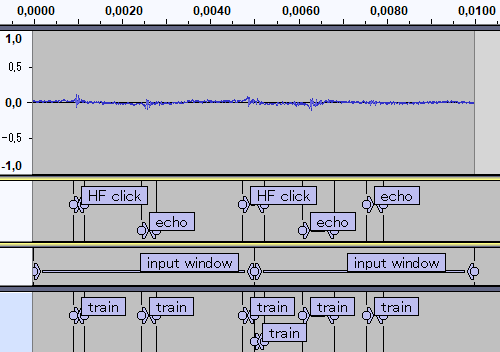}
\caption{A depiction on how multiple hand labeled annotations can exist in one input window, and how a hand labeled annotation can be split into multiple train entries in Audacity. On the top is the waveform of the two 5 millisecond input windows. The first text track displays the hand labeled annotations. The second text track shows the input windows. The last text track shows the train entries.}
\label{fig:traineventSplitMultiple}
\end{figure}

An annotation may be split across multiple input windows, which results in the number of annotation training entries being larger than the 6994 hand-labeled annotations. However, multiple annotation entries can exist within a single input window, this can be seen in Figure \ref{fig:traineventSplitMultiple}. As a result, only 5479 of the 38,405 files contain at least one annotation entry, which is fewer than the 6994 hand-labeled annotations. The remaining 32,926 input windows are empty, containing no annotation entries or YOLO bounding boxes.

\begin{figure}[!htb] \centering
\includegraphics[width=0.9\linewidth]{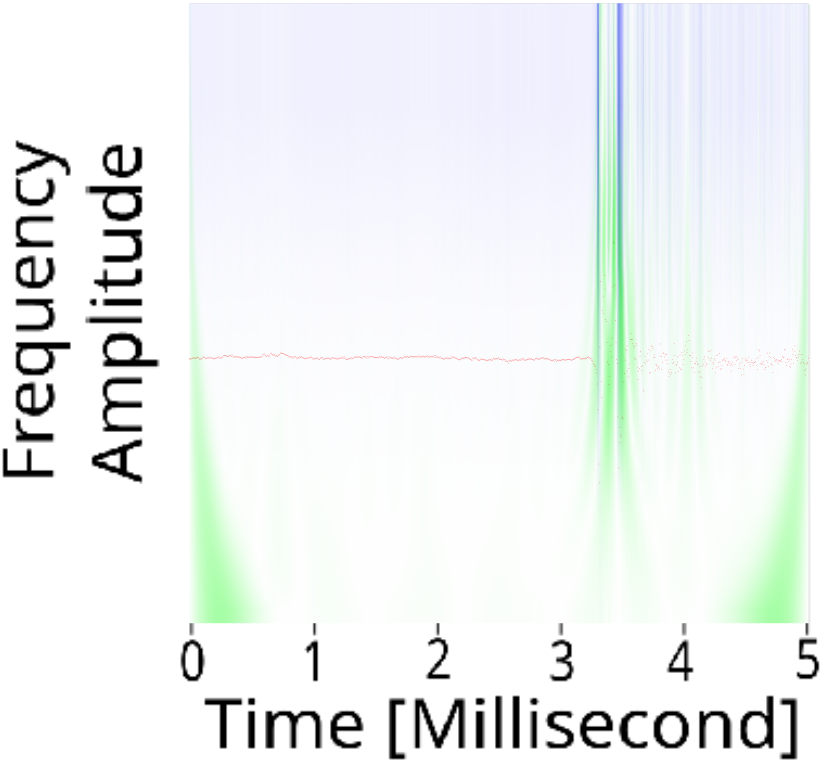}
\caption{Example of an SCWTSPEC input image. The image has dimensions of 960x960 pixels, with the red channel containing the waveform, the green channel containing the continuous wavelet transform (CWT), and the blue channel containing the spectrogram.}
\label{fig:SCWTSPEC}
\end{figure}

To prepare the data for network training, each window was preprocessed into a square 960x960x3 image representation, since the default YOLO uses square input images. The image representations were created using the waveform, spectrogram, and continuous wavelet transform (CWT). Specifically, the waveform, CWT, and spectrogram were encoded into the RGB channels of the resulting images. An example of these images, referred to as SCWTSPEC images, is shown in Figure \ref{fig:SCWTSPEC}. Although these images are challenging to interpret visually for humans, the color channels provide distinct features that are distinguishable for the network.
For training the networks, the dataset was split into three subsets: a training set consisting of 70\% of the samples (26,883), a validation set with 15\% of the samples (5761), and a test set containing the remaining 15\% (5761). The samples were randomly assigned to these subsets.
To evaluate the experiments, an additional 2 minute and 23 second file was provided with 5063 hand labeled event annotations (112 LF, 2617 HF, 3 US, 2732 clicks and 2331 echoes). During the development, it was found that the first hand labeled annotations were missing possible entries. As such, the hand annotations were expanded to a new improved version containing 5322 hand annotations (115 LF, 2742 HF, 6 US, 2863 clicks and 2459 echoes), as can be seen in Table \ref{tab:LFHFUSECHO}.
\section{Experiments and Results}
This chapter presents the experiments conducted to explore and validate the key concepts introduced in this thesis. 
The discussion follows a logical progression, beginning with the initial approaches tested and advancing through subsequent iterations. 
For each experiment, the results and shortcomings of the methods are summarized, and the rationale behind the transition to the next approach is provided. 
This structure highlights the iterative nature of the research, emphasizing how each experiment influenced the development of the next step in the process.

\subsection{PAMGuard Standalone Experiments}
To obtain comparable results, the first annotated data set with 6994 annotations was processed using PAMGuard and its built-in click detector. The experiment was conducted using the default click detector settings to assess how accurately PAMGuard could annotate the data based on the adjustable decibel threshold. Unfortunately, PAMGuard does not offer a direct method for converting click detections into Audacity annotations, so a small plugin was written to generate this conversion.
\begin{table}[!htb] \centering
    \begin{tabular}{|l|l|l|l|l|l|l|l|l|}
    \hline
        ~ & 9dB & 10dB & 13dB & 15dB & 17dB & 20dB & 30dB & 40dB  \\ \hline
        Detection & 47786 & 38887 & 19902 & 13607 & 9442 & 5673 & 672 & 103  \\ \hline
        TP & 13686 & 12650 & 9776 & 8185 & 6416 & 4354 & 617 & 96  \\ \hline
        FP & 34100 & 26237 & 10126 & 5422 & 3026 & 1319 & 55 & 7  \\ \hline
        Precision & 28.64 & 32.53 & 49.12 & 60.15 & 67.95 & 76.74 & 91.81 & 93.20  \\ \hline
        All & 6994 & 6994 & 6994 & 6994 & 6994 & 6994 & 6994 & 6994  \\ \hline
        Found & 6472 & 6218 & 5419 & 4617 & 3799 & 2709 & 437 & 79  \\ \hline
        Missed & 522 & 776 & 1575 & 2377 & 3195 & 4285 & 6557 & 6915  \\ \hline
        Recall & 92.53 & 88.90 & 77.48 & 66.01 & 54.31 & 38.73 & 6.24 & 1.12  \\ \hline
    \end{tabular}
    \caption{Results of the PAMGuard click detector on the train data with different decibel(dB) thresholds. The sum of all detections is the number of all PAMGuard detections using the threshold. The number of true positives (TP) is the number of PAMGuard detections that were within an event annotation. The number of false positives (FP) is the number of PAMGuard detections outside an event annotation. Precision is the percentage of correct predictions over all predictions. All is the number of annotations. The number of Found annotations is the number of event annotations that were found by at least one PAMGuard click detection. The number of Missed annotations is the number of annotations that have no PAMGuard click detection. The Recall is the percentage of found annotations over all annotations. The best overall accuracy was achieved with the 15 dB threshold, where PAMGuard identified 66.0\% of annotations with an accuracy of 60.2\%, resulting in an overall annotation accuracy of 39.7\%. The experiment was performed on the first data set with 6994 annotations later used for training. }
    \label{array:PAMres}
\end{table}
The results of this experiment are summarized in Table \ref{array:PAMres}. At lower decibel thresholds, PAMGuard successfully identifies many of the annotations, but it also generates a significantly higher number of false positives than true positives. As the decibel threshold increases, false positives decrease, but the overall number of detected events also drops. The optimal performance was achieved with the 15 dB threshold, where PAMGuard identified 66.0\% of annotations with an accuracy of 60.2\%, resulting in an overall annotation accuracy of 39.7\%. Due to the low overall accuracy and technical issues during development, experiments on the improved datasets or experiments to differentiate between clicks and echoes were not conducted on PAMGuard to save time.

\subsection{FOD only Experiments}
The FOD process for Dirac-like impulse detection, illustrated in Figure \ref{fig:FOD_Progression}, can itself be utilized as an event detection algorithm. To evaluate its performance, an experiment was conducted to compare the accuracy of FOD detection with that of machine learning models.
\begin{table}[!htb]
    \centering
        \begin{tabular}{|l|l|}
    \hline
        ~ & FOD  \\ \hline \hline
        Detection & 4834  \\ \hline
        TP & 4231  \\ \hline
        FP & 603  \\ \hline
        Precision & 87.52  \\ \hline \hline
        All & 5322  \\ \hline
        Found & 3232  \\ \hline
        Missed & 2090  \\ \hline
        Recall & 60.72  \\ \hline
    \end{tabular}
    \caption{Summary of results from the FOD-only mathematical approach. The Detection row represents the total number of FOD detections. True Positives (TP) are the FOD detections that correctly match an event annotation, while False Positives (FP) are detections that do not match any annotation. Precision is the proportion of correct predictions (TP) over total detections. All refers to the total number of annotations, with Found indicating annotations that were detected by at least one FOD, and Missed representing annotations with no FOD detections. Recall measures the proportion of found annotations over all annotations. Overall, 60.7\% of annotations were correctly found with an accuracy of 87.5\%, resulting in a final precision of 53.1\%. The experiment was performed on the later improved dataset with 5322 event annotations.
    }
    \label{array:FODres}
\end{table}
To summarize, the FOD detection method was able to find 87.5\% of annotated events with an annotation accuracy of 60.7\%. The overall accuracy of the FOD impulse detection was 53.1\%, which is a noticeable improvement over the PAMGuard click detections of 39.7\%. 
This experiment was performed at the same time the FOD box splicing solutions described in chapter \ref{Box_Slicing_Solutions} were added to CLICK-SPOT beta. As such, the experiment was performed on the later improved dataset with 5322 annotations.

\subsection{ANIMAL-SPOT Experiments}
The next step in this research was a proof-of-concept experiment using the ANIMAL-SPOT model to evaluate whether deep learning could be effectively applied to enhance click detection. 
The ANIMAL-SPOT model was initially trained using 20ms windows for binary event detection to determine whether the network could learn to perform the task with the limited dataset. 
The network settings used for this experiment are detailed in Table \ref{tab:ANIMAL-SPOTsettings}.
\begin{table}[!ht] \centering
    \begin{tabular}{|l|l|}
    \hline
        window size & 20ms/4ms \\ \hline
        window hop & 10ms/2ms \\ \hline
        lr & 10e-5 \\ \hline
        beta1 & 0.5 \\ \hline
        lr patience & 8 \\ \hline
        lr decay & 0.5 \\ \hline
        early stopping & 20 \\ \hline
        batch size & 16 \\ \hline
        n freq bins & 256 \\ \hline
        n fft & 128 \\ \hline
        hop & 32 \\ \hline
        kernel size & 7 \\ \hline
        sampling rate & 192,000 \\ \hline
        fmin & 2000 \\ \hline
        fmax & 90,000 \\ \hline
        augmentation & true \\ \hline
        min max norm & true \\ \hline
    \end{tabular}
    \caption{The settings used to train the ANIMAL-SPOT model. The window size specifies the size of the input images in milliseconds. The window hop represents the advancement time between consecutive windows in milliseconds. The learning rate (lr) is the preset learning rate. The adam optimizer (Beta1) controls the exponential decay rate of the first moment. Learning rate patience (lr patience) is the number of epochs without improvement on the validation set before the learning rate starts decaying. The learning rate decay (lr decay) factor determines the decay applied after the specified patience. Early stopping (early stopping) defines the number of epochs after which training stops if no improvement is observed, to prevent overfitting. Batch size is the number of images in each batch. The number of frequency bins (n freq bins) is the number of bins used to represent the given frequency range, from fmin to fmax. Number of FFT points (n fft) refers to the FFT window size in samples, and hop is the FFT hop size. The convolutional kernel size is the size of the initial square convolution in the ResNet architecture. Sampling rate is the rate at which the input signal is sampled. fmin and fmax represent the lower and upper-frequency thresholds for the ANIMAL-SPOT input image, respectively. Augmentation and min-max normalization were applied during training.  }
    \label{tab:ANIMAL-SPOTsettings}
\end{table}

\begin{figure}[!htb] \centering
\includegraphics[width=0.9\linewidth]{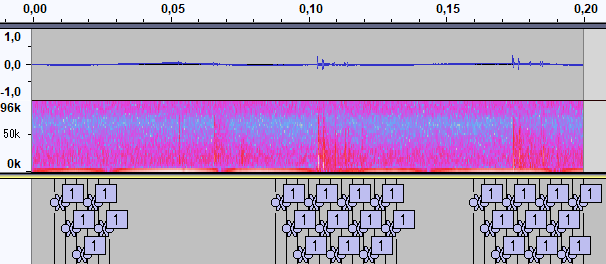}
\caption{Depiction of the ANIMAL-SPOT 20ms results in Audacity. Overall, the network was able to learn how to differentiate events from noise, but due to the large window size, the results are not usable for further progression. This depiction displays the 20ms ANIMAL-SPOT window results in a lower click density situation.}
\label{fig:AS20Excerpt}
\end{figure}
\begin{figure}[!htb] \centering
\includegraphics[width=0.9\linewidth]{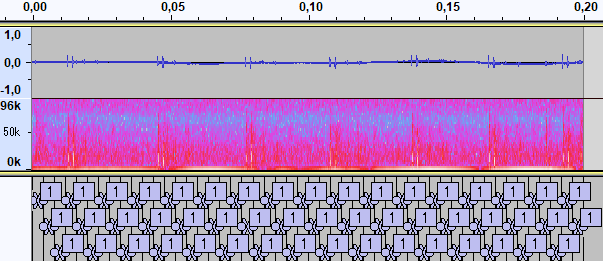}
\caption{Depiction of the ANIMAL-SPOT 20ms blocking problem in Audacity. The windows all merge into one large event, which makes it difficult to differentiate multiple events from each other. }
\label{fig:AS20Block}
\end{figure}
While the proof-of-concept experiment was successful in demonstrating that deep learning could be applied to click detection as seen in Figures \ref{fig:AS20Excerpt} and \ref{fig:AS20Block}, the results were ultimately not suitable for further experimentation. 
The task required the detection of individual events, but the 20ms overlapping windows were too large for accurate single-event binary classification. 
As a result, multiple events were too often grouped together within the same window, as shown in Image \ref{fig:AS20Block}, making it impossible to isolate individual events. 
To save on time, no further analysis or experiments were performed using the 20ms windows model, since the approach was flawed fundamentally. 
Instead, a follow-up experiment was conducted, where the window size was reduced from 20ms to 4ms.
This experiment aimed to test whether the ANIMAL-SPOT model could be adapted for single-event detection. 
The network settings for the 4ms window experiment are outlined in Table \ref{tab:ANIMAL-SPOTsettings}.

\begin{table}[!ht]
    \centering
\begin{tabular}{|l|l|l|l|l|l|l|l|l|l|}
    \hline
         ~ & 0.10 & 0.20 & 0.30 & 0.40 & 0.50 & 0.60 & 0.70 & 0.80 & 0.90  \\ \hline \hline
         Detection & 32294 & 24483 & 19918 & 16572 & 13885 & 11636 & 9652 & 7662 & 5520  \\ \hline
         TP & 11411 & 10814 & 10240 & 9679 & 9065 & 8364 & 7638 & 6620 & 5155  \\ \hline
         FP & 20883 & 13669 & 9678 & 6893 & 4820 & 3272 & 2014 & 1042 & 365  \\ \hline
         Precision & 35.33 & 44.16 & 51.41 & 58.40 & 65.28 & 71.88 & 79.13 & 86.40 & 93.38  \\ \hline\hline
         All & 5322 & 5322 & 5322 & 5322 & 5322 & 5322 & 5322 & 5322 & 5322  \\ \hline
         Found & 5149 & 5011 & 4903 & 4799 & 4685 & 4505 & 4288 & 3936 & 3316  \\ \hline
         Missed & 173 & 311 & 419 & 523 & 637 & 817 & 1034 & 1386 & 2006  \\ \hline
         Recall & 96.74 & 94.15 & 92.12 & 90.17 & 88.03 & 84.64 & 80.57 & 73.95 & 62.30 \\ \hline
    \end{tabular}
    \caption{This table presents the results of the ANIMAL-SPOT 4ms approach at varying thresholds from 0.1 to 0.9. The TP (true positives) row shows the number of ANIMAL-SPOT windows that correspond to an event, and the FP (false positives) row indicate the number of ANIMAL-SPOT windows that detected events without annotations. Finally, the Precision row provides the detection accuracy at each threshold.
    The Detection row shows the total number of events for each threshold, while the Found row indicate the number of events detected by ANIMAL-SPOT. The Missed row represents the number of events that were missed by ANIMAL-SPOT. The Recall measures the proportion of found annotations over all annotations. Overall, ANIMAL-SPOT achieved its best accuracy of 63.9\% (86.4\% label accuracy and 73.9\% detection accuracy) with a threshold of 80. The experiment was performed on the improved evaluation file with 5322 annotations. }
    \label{array:Animal-Spotres}
\end{table}

\begin{figure}[!htb] \centering
\includegraphics[width=0.9\linewidth]{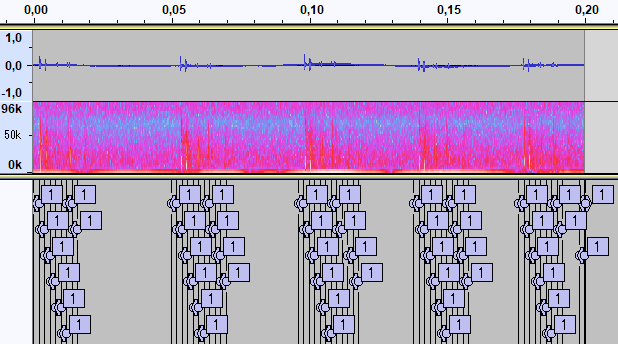}
\caption{Depiction of the ANIMAL-SPOT 4ms results in Audacity. Overall, the 4ms results are more precise than the 20ms results. Yet, the windows are still too large to differentiate between clicks and echoes. This depiction displays the 4ms ANIMAL-SPOT window results in a lower click density situation.}
\label{fig:AS4Excerpt}
\end{figure}
\begin{figure}[!htb] \centering
\includegraphics[width=0.9\linewidth]{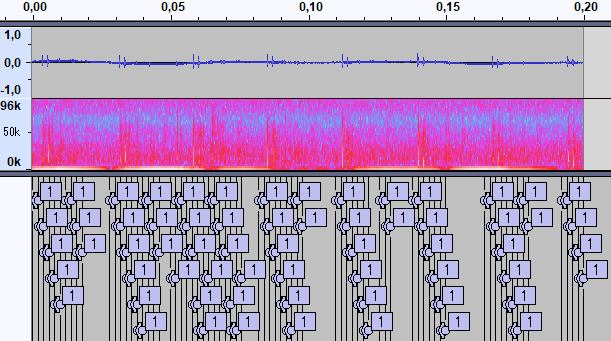}
\caption{This image was made in Audacity. Unlike the 20ms windows, the 4ms windows do not block into one event, so while it is better in differentiating events, it still blocks clicks and echo together. This depiction displays the 4ms ANIMAL-SPOT window results in a higher click density situation.}
\label{fig:AS4Block}
\end{figure}
Overall, the 4ms experiment achieved its best accuracy of 63.9\% (86.4\% label accuracy and 73.9\% detection accuracy) with a threshold of 80. This is a major improvement over the PAMGuard and FOD only approach.
Despite these adjustments, the results of the 4ms experiment revealed that while it was possible to train the network to detect clicks and echoes, the ANIMAL-SPOT model's binary classification continued to group multiple windows into one block, as can be seen in Figures \ref{fig:AS4Excerpt} and \ref{fig:AS4Block}. 
Additionally, isolated blocks containing single events were still too large to precisely define the single events. 
While the 4ms windows improved on the 20ms windows in burst, as can be seen when comparing Figure \ref{fig:AS20Block} and \ref{fig:AS4Block}, the 4ms windows are still too large. 
Creating smaller windows is difficult as the number of samples gets too small to generate a good image depiction.
These limitations indicated that the ANIMAL-SPOT model, despite its potential, was not suitable for the level of precision required for this task. 
Based on the insights gained from the ANIMAL-SPOT experiments, it became clear that a new approach, involving a different network architecture, would be necessary for subsequent experiments.

\subsection{YOLO Experiments}
After initial experimentation, the medium size YOLOv8 network model (YOLOv8m) was used as the baseline for the event detection task. 
In the first experiment, a 20ms time window was converted into a 416x416 spectral image, similar to the preprocessing method used for the ANIMAL-SPOT dataset. 
However, this approach failed as the YOLO model struggled to converge on this image representation. 
This prompted a reevaluation of the image conversion strategy.
Unlike the 4ms input window used in ANIMAL-SPOT, the YOLO network employs a 5ms window. With a sampling rate of 192,000 samples per second, this 5ms window corresponds to 960 samples. This adjustment was made to give the YOLO network more flexibility in handling multiple bounding boxes within a single input window, while also reducing the number of annotations split across windows. Additionally, it allows for a direct conversion of the 5ms window into a 960x960 pixel image without compression, as detailed in Chapter \ref{data}.
This direct conversion has several advantages. 
First, the sample count and image width are now consistent, enabling the use of various image representations, such as the continuous wavelet transform (CWT) and signal representations. 
Moreover, since all three conversion methods—signal, spectrogram, and CWT—generate single-channel grayscale images of the same size, these images can be combined into the red, green, and blue (RGB) channels of a standard color image. 
A fourth representation could theoretically be encoded into the alpha channel, but this approach was not explored in the context of this work.
The first experiment aimed to identify the optimal combination of conversion methods. 
The hypothesis was that if a particular conversion method did not contribute to improving model predictions, its corresponding channel could be removed, thereby reducing model size and enhancing performance. 
The results of this experiment are summarized in Table \ref{tab:ImageRepresentationExperiment}.

\begin{table}[!ht] 
    \resizebox{\columnwidth}{!}{
    \centering
    \begin{tabular}{|l|l|l|l|l|l|l|l|}
    \hline
        ~ & SIGNAL & CWT & SPEC & SCWT & SSPEC & CWTSPEC & SCWTSPEC  \\ \hline \hline
        Detection & 3545 & 5982 & 8678 & 5168 & 8359 & 5487 & 6255  \\ \hline
        TP Partial & 3118 & 4433 & 4970 & 3997 & 4894 & 4221 & 4675  \\ \hline
        TP Full & 2120 & 3998 & 4352 & 3478 & 4294 & 3697 & 4110  \\ \hline
        TP Partial\% & 87.95 & 74.11 & 57.27 & 77.34 & 58.54 & 76.92 & 74.74  \\ \hline
        TP Full\% & 59.80 & 66.83 & 50.14 & 67.29 & 51.36 & 67.37 & 65.70  \\ \hline
        FP & 427 & 1549 & 3708 & 1171 & 3465 & 1266 & 1580  \\ \hline
        FP\% & 12.04 & 25.89 & 42.72 & 22.65 & 41.45 & 23.07 & 25.25  \\ \hline \hline
        All & 5063 & 5063 & 5063 & 5063 & 5063 & 5063 & 5063  \\ \hline
        Partial & 3164 & 4656 & 4652 & 4329 & 4615 & 4470 & 4761  \\ \hline
        Full & 2035 & 4195 & 3983 & 3700 & 3997 & 3897 & 4269  \\ \hline
        Partial\% & 62.49 & 91.96 & 91.88 & 85.50 & 91.15 & 88.28 & 94.03  \\ \hline
        Full\% & 40.19 & 82.85 & 78.66 & 73.07 & 78.94 & 76.97 & 84.31  \\ \hline
        Missed & 1899 & 407 & 411 & 734 & 448 & 593 & 302  \\ \hline
        Missed\% & 37.50 & 8.03 & 8.11 & 14.49 & 8.84 & 11.71 & 5.96  \\ \hline
    \end{tabular}
        }
    \caption{Results of the representation combination experiment. The columns represent different combinations of image representations: SIGNAL (signal representation), CWT (continuous wavelet transformation), SPEC (spectrogram), SCWT (signal and continuous wavelet), SSPEC (signal and spectrogram), CWTSPEC (continuous wavelet transformation and spectrogram), and SCWTSPEC (all three representations combined). The Detection row indicates the number of detected events, while TP Partial and TP Full describe the true positive detections with overlaps of more than 20\% and 90\% with the hand annotated events, respectively. TP Partial\% and TP Full\% show the percentage of partial and full detections over all detections. FP describes the number of false positives, detected windows that have an overlap below 20\% or no overlap. FP\% is the percentage of false positives over all detection. All describes the number of all hand annotated events, Partial and Full describe the hand annotated events with overlaps of more than 20\% and 90\% with the detections. Partial\% and Full\% show the percentage of annotated events with partial and full overlaps over all annotated events. Missed describes the number of annotations with less than 20\% detection overlap, or no overlap. Missed\% is the percentage of missed annotations over all annotations. SCWTSPEC achieved the best partial and full overlap percentages for ground truth labels. This proves that every representation adds a noticeable improvement to the model. This early experiment was performed on the first evaluation dataset with 5063 event annotations. An experiment on the improved dataset was deemed unnecessary.}
    \label{tab:ImageRepresentationExperiment}
\end{table}

\subsubsection{Confidence Threshold Experiments}
\begin{figure}[!htb] \centering
\includegraphics[width=0.9\linewidth]{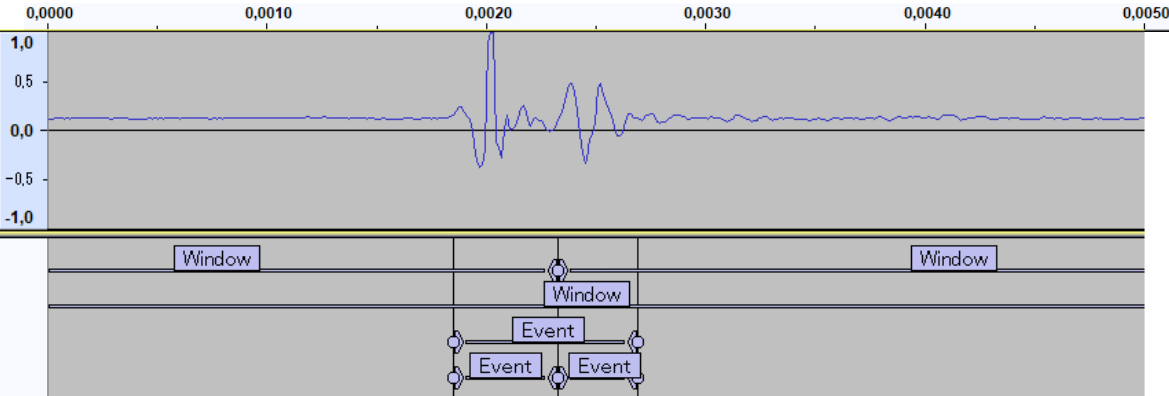}
\caption{Example depiction of a split event. The windows end and start in the middle of the
click, as such three events were generated instead of one.}
\label{fig:MultipleBoxesPerEvent}
\end{figure}

The next experiment focused on identifying the most suitable confidence threshold for the event analysis task. The goal was to find a threshold that maximized recall, ensuring that as many clicks as possible were detected, even if it led to a higher false positive rate. The results of this confidence threshold experiment are detailed in Table \ref{tab:YOLOConfRes}. One challenge encountered during this experiment was the model generating multiple bounding boxes for the same event, due to the hop size of 2.5ms. This issue could result in up to three boxes for a single event, especially if the event was split, as can be seen in Figure \ref{fig:MultipleBoxesPerEvent}. Additionally, events occurring within 2 milliseconds of each other often resulted in overlapping boxes. Since overlapping events were not desired, a solution was
needed to merge these boxes. The first approach involved a simple full merge, where any overlapping bounding boxes were combined into a single larger box. While this eliminated the overlapping boxes, it introduced a new issue: the merged boxes often contained multiple events, leading to misclassifications. To address this, two box-slicing methods were tested.
\begin{table}[!htb]
    \centering
    \begin{tabular}{|l|l|l|l|l|l|l|}
    \hline
        ~ & 13 & 14 & 15 & 16 & 17 & 19  \\ \hline \hline
        Detection & 4487 & 4491 & 4515 & 4520 & 4496 & 4504  \\ \hline
        TP Partial & 3018 & 3068 & 3115 & 3183 & 3250 & 3337  \\ \hline
        TP Full & 2723 & 2753 & 2780 & 2835 & 2876 & 2916  \\ \hline
        TP Partial\% & 67.26 & 68.31 & 68.99 & 70.42 & 72.28 & 74.08  \\ \hline
        TP Full\% & 60.68 & 61.30 & 61.57 & 62.72 & 63.96 & 64.74  \\ \hline
        FP & 1469 & 1423 & 1400 & 1337 & 1246 & 1167  \\ \hline
        FP\% & 32.73 & 31.68 & 31.00 & 29.57 & 27.71 & 25.91  \\ \hline \hline
        All & 5063 & 5063 & 5063 & 5063 & 5063 & 5063  \\ \hline
        Partial & 4794 & 4781 & 4761 & 4749 & 4730 & 4707  \\ \hline
        Full & 4386 & 4346 & 4306 & 4268 & 4216 & 4138  \\ \hline
        Partial\% & 94.68 & 94.43 & 94.03 & 93.79 & 93.42 & 92.96  \\ \hline
        Full\% & 86.62 & 85.83 & 85.04 & 84.29 & 83.27 & 81.73  \\ \hline
        Missed & 269 & 282 & 302 & 314 & 333 & 356  \\ \hline
        Missed\% & 5.31 & 5.56 & 5.9 & 6.2 & 6.57 & 7.03  \\ \hline
    \end{tabular}
    \caption{Results of the confidence threshold experiment. The table shows the performance of the YOLO model at different confidence threshold values (ranging from 13 to 19 from a confidence between 1 and 100) for event detection. The results are portioned the same as in Table \ref{tab:ImageRepresentationExperiment}. The experiment demonstrates that adjusting the confidence threshold serves as a tradeoff between false positives and the number of detected events, with higher thresholds increasing partial overlaps in both YOLO detections and ground truth events. This experiment was also performed on the first evaluation dataset with 5063 event annotations instead of the later improved dataset with 5322. A redo of this experiment was deemed unnecessary.}
    \label{tab:YOLOConfRes}
\end{table}

\subsubsection{FOD post-processing to enhance bounding box position}
The first slicing method attempted to divide the merged boxes based on confidence values and overlap of the bounding boxes, aiming to determine which parts of a box belonged to the same event. However, this approach was not effective when the event was split, leading
to a consistent problem where two bounding boxes were part of the same event with no overlap, which was deemed too difficult to eliminate based on the approach. The second method involved using the first order detection (FOD) technique to accurately define the time frames for events, allowing more precise splicing decisions based on confidence levels and inter-arrival times. The process of this approach can be seen in Figure \ref{fig:FOD_Progression}. The results of the FOD-based splicing method are presented in Table \ref{tab:FOD_box_solution2}. In addition, a new adjustable confidence threshold based on the FOD was added to reduce the number of false positives, while this new addition slightly decreased the overall findings it improved the false positive rate remarkably. This method successfully enabled the model to output single-event windows, resolving the issue of overlapping events.

\begin{table}[!ht]
    \centering
    \begin{tabular}{|l|l|}
    \hline
        Detection & 6416  \\ \hline
        TP Partial & 5103  \\ \hline
        TP Full & 2967  \\ \hline
        TP Partial\% & 79.53  \\ \hline
        TP Full\% & 46.24  \\ \hline
        FP & 1313  \\ \hline
        FP\% & 20.46  \\ \hline
        All & 5322  \\ \hline
        Partial & 4780  \\ \hline
        Full & 2628  \\ \hline
        Partial\% & 89.81  \\ \hline
        Full\% & 49.37  \\ \hline
        Missed & 542  \\ \hline
        Missed\% & 10.18  \\ \hline
    \end{tabular}
    \caption{The experiment was performed on the improved evaluation data. As such, there are 5322 instead of 5063 Ground truth labels. The bounding boxes were provided with a confidence threshold of 0.15 (see Table \ref{tab:YOLOConfRes}). The results are partitioned the same as in Table \ref{tab:ImageRepresentationExperiment}. The FOD splitting increased the number of events from 4487 to 6416. In addition, the split bounding boxes were less likely to have a full overlap. Due to the FOD confidence threshold adjustment, the overall findings of the labels were partially reduced from 94.0\% to 89.8\%, but the accuracy of the detector rose from 69.0\% to 79.5\% leading to a reduction of false positives. These changes increased the overall accuracy from 63.0\% to 71.4\%.}
    \label{tab:FOD_box_solution2}
\end{table}

\subsection{Random Forest Click and Echo Differentiation}
Despite successfully identifying single-event windows, the model still struggled to differentiate between clicks and echoes. 
This challenge arises from the similarity in the intensity, reverberation, and duration of click signals, which makes it difficult to distinguish clicks from echoes when analyzed in isolation. 
However, when clicks and echoes are observed in proximity and considered as pairs, differentiation becomes more feasible.
To address this issue, a new post-processing method was developed to distinguish clicks, echoes and misclassifications (grouped as other) within the identified single-event windows. 
This method leverages the spatial and temporal relationships between detected events, enhancing the accuracy of click-echo classification.
The chosen approach involved using key defining features and introducing a random forest classifier for click and other differentiation. 
The input vector values for the random forest classifier are provided in Table \ref{tab:RamdomForest}. 
\begin{table}[!htb] \centering
    \begin{tabular}{|l|}
    \hline
    Random Forest Classifier input vector \\ \hline \hline
        start  \\ \hline
        end  \\ \hline
        confidence \\ \hline
        length  \\ \hline
        number FOD  \\ \hline
        minimum energy  \\ \hline
        maximum energy  \\ \hline
        mean energy  \\ \hline
        max FOD  \\ \hline
        FOD direction  \\ \hline
        strongest frequency  \\ \hline
        interarrival  \\ \hline
    \end{tabular}
        \caption{The start and end represent the event’s beginning and end times. The confidence represents the confidence of the YOLO detection for the bounding box. The length describes the event's duration, and the number FOD refers to the number of FOD peaks within the bounding box. Additional features include the minimum and maximum energy levels, the mean energy, and the max FOD peak, which represents the strongest FOD peak. The FOD direction reflects the phase of the strongest FOD peak, and the strongest frequency identifies the frequency bin with the highest energy. Finally, the interarrival time measures the gap between the current and next bounding box.}
    \label{tab:RamdomForest}
\end{table}
Three experimental approaches were tested. 
Ten binary trees were constructed to test for consistency.
The first experiment used three events, one from the past, the current one and one from the future, to differentiate clicks from others. 
The best accuracy achieved was 85.3\%, which was promising but left room for improvement.
The second experiment employed five events, with four from the past and the current one. 
This method operated under the assumption that buffering window events to access future events was unnecessary. 
This approach yielded a significant improvement, with the best accuracy reaching 90.4\%.
The third experiment incorporated nine events, with four from the past, the current one and four from the future. 
The best of this method achieved an accuracy of 89.1\%, which is comparable to the four from the past model. Since it did not offer significant improvements over the five-event method, the five-event approach was selected for click and other differentiation.
When disregarding the echo and misclassification labels, the best click label accuracy that the random forest classifier achieved was 95.93\%. 

\subsection{Final Optimizations}
Until this point, recall was the primary metric for optimizing detection. 
This meant that the model aimed to identify as many clicks as possible, accepting a reasonable number of false positives. 
Table \ref{tab:OPTIMIZATION1} and Table \ref{tab:OPTIMIZATION2} show that the model with a lower confidence threshold was most effective for this purpose.
In a subsequent meeting with Dr. Vester, it was decided that the model's results should operate independently, with a focus on the clicks themselves. 
The goal was for the model to function without human supervision or the need for corrections. 
As a result, the optimization shifted towards maximizing overall accuracy rather than recall. 
While this decision inevitably led to a reduction in recall and, consequently, the number of calls detected, it also reduced the occurrence of false positives. 
To assess this change, the confidence threshold was reevaluated. 
The results of this accuracy focused experiment are presented in Tables \ref{tab:OPTIMIZATION1} and \ref{tab:OPTIMIZATION2}.
\begin{table}[!ht] \centering
    \resizebox{0.75\columnwidth}{!}{
   \begin{tabular}{|l|l|l|l|l|l|l|l|}
    \hline
        ~ & confidence & 15 & 20 & 25 & 30 & 35 & 40  \\ \hline \hline
        YOLO detection & Detection & 6416 & 6070 & 5793 & 5559 & 5360 & 5181  \\ \hline
        ~ & TP Partial & 5103 & 4947 & 4809 & 4691 & 4571 & 4472  \\ \hline
        ~ & TP Full & 2967 & 2861 & 2770 & 2688 & 2608 & 2544  \\ \hline
        ~ & TP Partial\% & 79.53 & 81.49 & 83.01 & 84.38 & 85.27 & 86.31  \\ \hline
        ~ & TP Full\% & 46.24 & 47.13 & 47.81 & 48.35 & 48.65 & 49.10  \\ \hline
        ~ & FP & 1313 & 1123 & 984 & 868 & 789 & 709  \\ \hline
        ~ & FP\% & 20.46 & 18.50 & 16.98 & 15.61 & 14.72 & 13.68  \\ \hline \hline
        YOLO click & Detection & 2988 & 2880 & 2793 & 2683 & 2707 & 2730  \\ \hline
        ~ & TP Partial & 2600 & 2586 & 2565 & 2574 & 2529 & 2496  \\ \hline
        ~ & TP Full & 1733 & 1687 & 1648 & 1629 & 1576 & 1538  \\ \hline
        ~ & TP Partial\% & 87.01 & 89.79 & 91.83 & 95.93 & 93.42 & 91.42  \\ \hline
        ~ & TP Full\% & 57.99 & 58.57 & 59.01 & 60.71 & 58.21 & 56.33  \\ \hline
        ~ & FP & 388 & 294 & 228 & 109 & 178 & 234  \\ \hline
        ~ & FP\% & 12.98 & 10.20 & 8.16 & 4.06 & 6.57 & 8.57  \\ \hline \hline
        YOLO echo & Detection & 3428 & 3190 & 3000 & 2876 & 2653 & 2451  \\ \hline
        ~ & TP Partial & 2207 & 2125 & 2042 & 1991 & 1901 & 1810  \\ \hline
        ~ & TP Full & 1060 & 1029 & 994 & 970 & 935 & 897  \\ \hline
        ~ & TP Partial\% & 64.38 & 66.61 & 68.06 & 69.22 & 71.65 & 73.84  \\ \hline
        ~ & TP Full\% & 30.92 & 32.25 & 33.13 & 33.72 & 35.24 & 36.59  \\ \hline
        ~ & FP & 1221 & 1065 & 958 & 885 & 752 & 641  \\ \hline
        ~ & FP\% & 35.61 & 33.38 & 31.93 & 30.77 & 28.34 & 26.15  \\ \hline \hline
        Hand Annotation & All & 5322 & 5322 & 5322 & 5322 & 5322 & 5322  \\ \hline
        ~ & Partial & 4780 & 4686 & 4598 & 4517 & 4437 & 4370  \\ \hline
        ~ & Full & 2628 & 2562 & 2508 & 2456 & 2401 & 2358  \\ \hline
        ~ & Partial\% & 89.81 & 88.04 & 86.39 & 84.87 & 83.37 & 82.11  \\ \hline
        ~ & Full\% & 49.37 & 48.13 & 47.12 & 46.14 & 45.11 & 44.30  \\ \hline
        ~ & Missed & 542 & 636 & 724 & 805 & 885 & 952  \\ \hline
        ~ & Missed\% & 10.18 & 11.95 & 13.60 & 15.12 & 16.62 & 17.88  \\ \hline \hline
        Annotated click & All & 2864 & 2864 & 2864 & 2864 & 2864 & 2864  \\ \hline
        ~ & Partial & 2535 & 2516 & 2497 & 2502 & 2469 & 2449  \\ \hline
        ~ & Full & 1704 & 1659 & 1623 & 1605 & 1556 & 1522  \\ \hline
        ~ & Partial\% & 88.51 & 87.84 & 87.18 & 87.36 & 86.20 & 85.50  \\ \hline
        ~ & Full\% & 59.49 & 57.92 & 56.66 & 56.04 & 54.32 & 53.14  \\ \hline
        ~ & Missed & 329 & 348 & 367 & 362 & 395 & 415  \\ \hline
        ~ & Missed\% & 11.48 & 12.15 & 12.81 & 12.63 & 13.79 & 14.49  \\ \hline \hline
        Annotated echo & All & 2458 & 2458 & 2458 & 2458 & 2458 & 2458  \\ \hline
        ~ & Partial & 1911 & 1867 & 1820 & 1796 & 1737 & 1669  \\ \hline
        ~ & Full & 835 & 826 & 818 & 819 & 806 & 785  \\ \hline
        ~ & Partial\% & 77.74 & 75.95 & 74.04 & 73.06 & 70.66 & 67.90  \\ \hline
        ~ & Full\% & 33.97 & 33.60 & 33.27 & 33.31 & 32.79 & 31.93  \\ \hline
        ~ & Missed & 547 & 591 & 638 & 662 & 721 & 789  \\ \hline
        ~ & Missed\% & 22.25 & 24.04 & 25.95 & 26.93 & 29.33 & 32.09  \\ \hline
    \end{tabular}
    }
    \caption{Results of the optimization experiments. The experiment was performed on the improved dataset with 5322 labeled events. The rows are described identical to Table \ref{tab:ImageRepresentationExperiment}. The confidence threshold were increased to see if a better recall can be achieved. Clicks and echoes were also separated to see how the model confidence threshold would change the overlap with the annotations. This table only shows the first half of the experiment, from the confidence threshold 15 to 40. The second half from 45 to 75 is displayed in Table \ref{tab:OPTIMIZATION2}. The 30 confidence threshold gave the best recall from the experiment. As such, a threshold of 30 was used for the rest of this study.}
    \label{tab:OPTIMIZATION1}
\end{table}

\begin{table}[!htb] \centering
    \resizebox{0.75\columnwidth}{!}{
    \begin{tabular}{|l|l|l|l|l|l|l|l|l|}
    \hline
        ~ & confidence & 45 & 50 & 55 & 60 & 65 & 70 & 75  \\ \hline \hline
        YOLO detection & Detection & 5024 & 4887 & 4770 & 4664 & 4556 & 4457 & 4362  \\ \hline
        ~ & TP Partial & 4391 & 4304 & 4224 & 4156 & 4082 & 4013 & 3953  \\ \hline
        ~ & TP Full & 2490 & 2438 & 2387 & 2326 & 2290 & 2262 & 2236  \\ \hline
        ~ & TP Partial\% & 87.40 & 88.07 & 88.55 & 89.10 & 89.59 & 90.03 & 90.62  \\ \hline
        ~ & TP Full\% & 49.56 & 49.88 & 50.04 & 49.87 & 50.26 & 50.75 & 51.26  \\ \hline
        ~ & FP & 633 & 583 & 546 & 508 & 474 & 444 & 409  \\ \hline \hline
        ~ & FP\% & 12.59 & 11.92 & 11.44 & 10.89 & 10.40 & 9.96 & 9.37  \\ \hline
        YOLO click & Detection & 2726 & 2735 & 2730 & 2714 & 2716 & 2691 & 2680  \\ \hline
        ~ & TP Partial & 2452 & 2424 & 2405 & 2378 & 2361 & 2334 & 2301  \\ \hline
        ~ & TP Full & 1505 & 1486 & 1468 & 1433 & 1423 & 1403 & 1381  \\ \hline
        ~ & TP Partial\% & 89.94 & 88.62 & 88.09 & 87.61 & 86.92 & 86.73 & 85.85  \\ \hline
        ~ & TP Full\% & 55.20 & 54.33 & 53.77 & 52.80 & 52.39 & 52.13 & 51.52  \\ \hline
        ~ & FP & 274 & 311 & 325 & 336 & 355 & 357 & 379  \\ \hline
        ~ & FP\% & 10.05 & 11.37 & 11.90 & 12.38 & 13.07 & 13.26 & 14.14  \\ \hline \hline
        YOLO echo & Detection & 2298 & 2152 & 2040 & 1950 & 1840 & 1766 & 1682  \\ \hline
        ~ & TP Partial & 1737 & 1665 & 1604 & 1563 & 1506 & 1459 & 1412  \\ \hline
        ~ & TP Full & 860 & 829 & 794 & 770 & 737 & 717 & 695  \\ \hline
        ~ & TP Partial\% & 75.58 & 77.36 & 78.62 & 80.15 & 81.84 & 82.61 & 83.94  \\ \hline
        ~ & TP Full\% & 37.42 & 38.52 & 38.92 & 39.48 & 40.05 & 40.60 & 41.31  \\ \hline
        ~ & FP & 561 & 487 & 436 & 387 & 334 & 307 & 270  \\ \hline
        ~ & FP\% & 24.41 & 22.63 & 21.37 & 19.84 & 18.15 & 17.38 & 16.05  \\ \hline \hline
        Hand Annotations & All & 5322 & 5322 & 5322 & 5322 & 5322 & 5322 & 5322  \\ \hline
        ~ & Partial & 4306 & 4245 & 4186 & 4125 & 4060 & 4001 & 3952  \\ \hline
        ~ & Full & 2321 & 2289 & 2259 & 2209 & 2185 & 2171 & 2156  \\ \hline
        ~ & Partial\% & 80.90 & 79.76 & 78.65 & 77.50 & 76.28 & 75.17 & 74.25  \\ \hline
        ~ & Full\% & 43.61 & 43.01 & 42.44 & 41.50 & 41.05 & 40.79 & 40.51  \\ \hline
        ~ & FP & 1016 & 1077 & 1136 & 1197 & 1262 & 1321 & 1370  \\ \hline
        ~ & FP\% & 19.09 & 20.23 & 21.34 & 22.49 & 23.71 & 24.82 & 25.74  \\ \hline \hline
        Annotated click & All & 2864 & 2864 & 2864 & 2864 & 2864 & 2864 & 2864  \\ \hline
        ~ & Partial & 2412 & 2390 & 2374 & 2351 & 2339 & 2313 & 2287  \\ \hline
        ~ & Full & 1491 & 1472 & 1455 & 1423 & 1415 & 1396 & 1377  \\ \hline
        ~ & Partial\% & 84.21 & 83.44 & 82.89 & 82.08 & 81.66 & 80.76 & 79.85  \\ \hline
        ~ & Full\% & 52.06 & 51.39 & 50.80 & 49.68 & 49.40 & 48.74 & 48.07  \\ \hline
        ~ & FP & 452 & 474 & 490 & 513 & 525 & 551 & 577  \\ \hline
        ~ & FP\% & 15.78 & 16.55 & 17.10 & 17.91 & 18.33 & 19.23 & 20.14  \\ \hline \hline
        Annotated echo & All & 2458 & 2458 & 2458 & 2458 & 2458 & 2458 & 2458  \\ \hline
        ~ & Partial & 1610 & 1550 & 1505 & 1472 & 1423 & 1385 & 1346  \\ \hline
        ~ & Full & 757 & 735 & 715 & 697 & 672 & 660 & 643  \\ \hline
        ~ & Partial\% & 65.50 & 63.05 & 61.22 & 59.88 & 57.89 & 56.34 & 54.75  \\ \hline
        ~ & Full\% & 30.79 & 29.90 & 29.08 & 28.35 & 27.33 & 26.85 & 26.15  \\ \hline
        ~ & FP & 848 & 908 & 953 & 986 & 1035 & 1073 & 1112  \\ \hline
        ~ & FP\% & 34.49 & 36.94 & 38.77 & 40.11 & 42.10 & 43.65 & 45.24  \\ \hline
    \end{tabular}
    }
    \caption{Second half of the experiment from Table \ref{tab:OPTIMIZATION1}. }
    \label{tab:OPTIMIZATION2}
\end{table}
Another key point of discussion was the emphasis on click rate rather than the sheer number of findings. 
For Dr. Vester and her team, the click rate would serve as an indicator of activity for individual animals or groups. 
Therefore, the updated approach did not require perfect click detection. Instead, it aimed for a strong correlation between the detected click rate and the actual click rate. To this end, the event rate, click rate and echo rate were calculated over the annotated evaluation file and the CLICK-SPOT final toolchain output using multiple thresholds. 
The outcomes of the click rate analysis are shown in Table \ref{tab:clickrate}.
\begin{table}[!ht] \centering
    \begin{tabular}{|l|l|l|l|}
    \hline
        Confidence & Event Correlation & Click correlation & Echo correlation \\ \hline
        15 & 94.51 & 97.84 & 84.87  \\ \hline
        20 & 94.79 & 98.24 & 85.20  \\ \hline
        25 & 95.04 & 98.37 & 85.86  \\ \hline
        30 & 95.06 & 98.48 & 84.36  \\ \hline
        35 & 95.03 & 98.00 & 85.37  \\ \hline
        40 & 94.92 & 98.10 & 86.34  \\ \hline
        45 & 94.94 & 97.57 & 86.29  \\ \hline
        50 & 94.95 & 97.32 & 86.18  \\ \hline
        55 & 94.69 & 97.02 & 85.84  \\ \hline
        60 & 94.60 & 96.75 & 86.39  \\ \hline
        65 & 94.34 & 96.52 & 86.62  \\ \hline
        70 & 94.20 & 96.51 & 85.93  \\ \hline
        75 & 93.92 & 95.92 & 86.24  \\ \hline
    \end{tabular}
    \caption{The event rate, click rate and echo rate correlation results based on the final CLICK-SPOT model confidence threshold. Overall, the correlation between the found click events and the annotated click events is higher than between the echoes. The best results were achieved with a confidence threshold of 30.}
    \label{tab:clickrate}
\end{table}

\section{Discussion}
In this chapter, the findings from the previous experiments are analyzed to highlight the key insights gained from the solutions, as well as the challenges that remain to be addressed in future research.

\subsection{PAMGuard}
The low overall accuracy of 39.7\% (see Table \ref{array:PAMres}) for PAMGuard can be attributed to the fact that its click detector was not designed for this specific task. Originally, the click detector was intended to be used in conjunction with the click-bearing localizer to determine the direction of incoming clicks, which is essential for tracking animals. The localizer does not require all clicks to track an animal, only the most prominent peak clicks. As a result, a higher decibel threshold for the click detector can serve as an effective preprocessor for the click-bearing calculator.
Additionally, the plugin that converted click detections into Audacity annotations was unstable. This instability lead to unexpected crashes during the experiment. The cause of this instability was not identified. To work around this, the input data was split into smaller 10-second segments, and failed experiments were rerun until successful.

\subsection{Standalone FOD event detection}
With an overall accuracy of 53.1\% (see Table \ref{array:FODres}) and no crashes, the FOD event detection performed better for the click annotation task compared to the PAMGuard click detector. However, the FOD detector alone produced a high number of false positives. Despite this, it proved effective for identifying peaks within YOLO bounding boxes and was therefore incorporated into the box-slicing solution to more accurately differentiate individual events.

\subsection{ANIMAL-SPOT}
The ANIMAL-SPOT model was originally designed for call classification using window segmentation. It is designed to find and classify longer calls in an audio file, usually in seconds, not milliseconds. Since most animals communicate vividly when socializing, it is not uncommon to have multiple overlapping calls from multiple animals.
While it proved effective for detecting calls with an overall accuracy of 63.9\% (86.4\% label accuracy and 73.9\% detection accuracy, see Table \ref{array:Animal-Spotres}) with a threshold of 80, ANIMAL-SPOT was not designed to isolate calls within a window. Overall, its segmentation approach was not precise enough for the millisecond-level events and short inter-arrival times involved in click and echo event detection. The problem was that the overlapping windows would block together, making single-event differentiation unfeasible. Given that ANIMAL-SPOT was not intended for this type of task, augmenting the model to meet these specific requirements would have been more complex than simply adopting a different model approach.
However, the training process with ANIMAL-SPOT provided valuable insights into the feasibility of single-event detection. Despite the limitations in segmentation, the network demonstrated that the training data could indeed be used to develop a model capable of detecting isolated events, laying the groundwork for the CLICK-SPOT toolchain.

\subsection{YOLO}
YOLO was deemed the most suitable solution for the event detection task. By transforming the bounding boxes into timestamps and applying a first-order detection method, the CLICK-SPOT beta isolated event detector model achieved an accuracy of 68.86\% (with an event detection precision of 74.08\% and a detection recall of 92.96\%, see Table \ref{tab:YOLOConfRes}). 
However, the main limitation of the YOLO approach was its treatment of each image as an isolated input, preventing the model from utilizing interconnected information between images. While YOLO excelled at detecting the Dirac-like clicks and echoes amidst surrounding noise, it struggled to differentiate between these events due to their high variability and similarities. 

\subsection{Random Forest}
There are several approaches available for differentiating clicks and echoes. In this work, the random forest classifier was selected due to its ease of training and implementation. Additionally, the task was expanded to identify potential misclassifications from YOLO and filter them out alongside the echoes. Overall, with a label accuracy of 71.42\% (with an event detection precision of 79.53\% and a detection recall of 89.81\%, see Table \ref{tab:FOD_box_solution2}), the random forest model proved effective when combined with YOLO, forming a robust toolchain for the task.

\subsection{Final Optimizations}
With the addition of the random forest, the final task in this work was to optimize the final CLICK-SPOT toolchain. After discussions with Dr. Vester, the goal for this network was to operate fully autonomously, meaning without human supervision. Combining all results, CLICK-SPOT achieves a click classification accuracy of 82.56\%, with a click detection precision of 86.32\% and a click label recall of 95.93\% (see Table \ref{tab:OPTIMIZATION1}). The detected clicks have a correlation of 98.01\% with the annotations (see Table \ref{tab:clickrate}). This is a major improvement when compared to the prior attempts.
While the primary objective of the click detector was to use its results as an indicator of activity, the error and accuracy rates were secondary to the correlation with observed data. As long as the detector's results strongly aligned with the seen results, the exact accuracy was less critical. The strong correlation indicates that the model’s error of 17.44\% is distributed relatively evenly across the results, meaning that despite the relatively high error rate compared to the correlation, the tool can still perform its intended function effectively.
Additionally, the network takes 25 minutes of real-time processing to analyze just 1 minute of material. With a real-time factor of 25, this means the tool is not yet effective for field use in its current iteration. However, the task of labeling a data corpus can be parallelized, since multiple CLICK-SPOT instances can work independently on multiple files, as can be seen in a similar approach by ANIMAL-SPOT during the ORCA-SLANG runtime experiments\cite{Bergler21}. In this work, ANIMAL-SPOT was used to process 20,000h of underwater recordings to remove noise. The ORCA-SLANG process was able to reduce the 20,000 hours (833 days) into 3000 hours (125 days) through parallel processing in 14 days. This process could also be done with a data stream of a passive observation hydrophone, but it would not work in a restrictive fieldwork environment where processing or power are limited. 

\subsection{Future Work} 
\begin{figure}[!htb] \centering
  \includegraphics[width=0.9\linewidth]{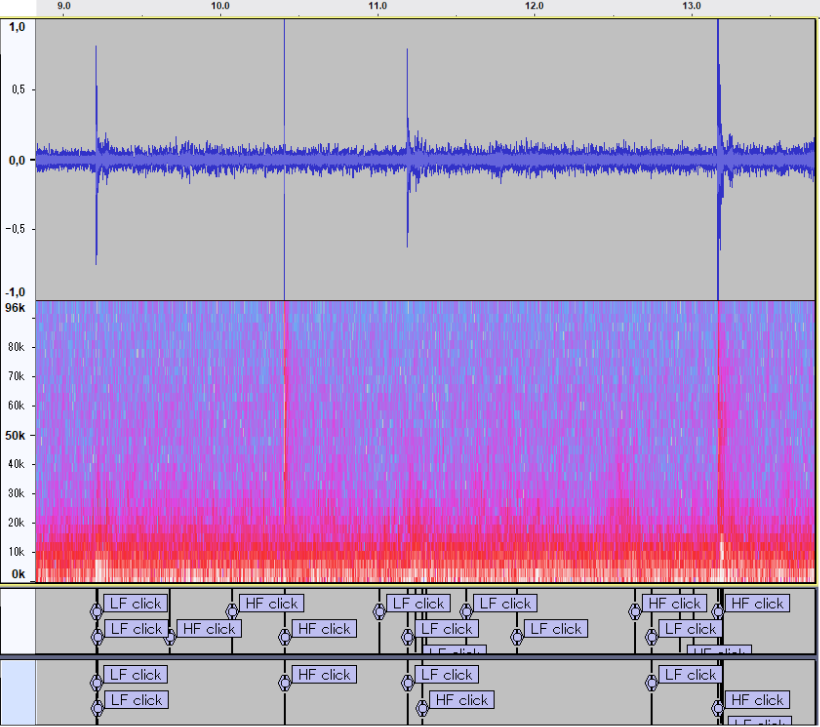}
  \caption{Extract of the sperm whale track in Audacity. The first text labels show the results of the unoptimized approach. The second text label shows the optimized approach.}
\label{fig:OtherAnimal}
\end{figure}

The development of the CLICK-SPOT toolchain has not only led to valuable insights but has also generated several promising directions for further ideas and additions to the toolchain. The most prominent enhancements for CLICK-SPOT are summarized below. 
To use the CLICK-SPOT toolchain in the field, the network has to be optimized for real-time processing. This would enable field deployment where immediate feedback is critical, significantly improving the tool's applicability in dynamic environments.
To that regard, the YOLO network could be integrated with contextual information. Combining object detection with context over multiple windows, like a recurrent neural network YOLO, could improve the detection accuracy of the toolchain and the inference speed of the model. It could also be used to possibly differentiate between clicks and echo directly, without the need of another context classifier. Another improvement could be the reduction of the input image. By combining the information of the 960x960x3 images into smaller dimensions, the model inference time would decrease as well. yet, even with all these possible improvements realized, it is unlikely that the model would have a real-time factor of 1. 
A weaker, but faster, detection system, like PAMGUARD, ANIMAL-SPOT, or a variation of the FOD detector could be used as a pre-detector to filter out areas of no interest. Through this, the CLICK-SPOT model would act as a detection validator to gain more precise time windows. These time windows could then be used by a following localizer algorithm to find and monitor the animal. \\
Another addition would be the differentiation between low-frequency (LF), high-frequency (HF), and ultrasonic-frequency (US) clicks. A future version of CLICK-SPOT could incorporate a classification system to automatically differentiate between various types of clicks based on the frequency ranges. The current final iteration of CLICK-SPOT already had an approach for click differentiation, but due to the high amount of noise in the lower frequency range and the small amount of available US clicks, the approach has difficulties differentiating between LF, HF and US clicks. Through methods, such as high pass filter and noise removal detraction, the differentiator achieved an accuracy of 77\%. While this is a good start, better methods could improve this accuracy in the future.
With that in mind, two more additions would be to add the click rate to the toolchain and to calculate and display the frequency of detected clicks over time. \\ 
While CLICK-SPOT was initially developed for click and echo differentiation in killer whales, the tool is adaptable to other species through retraining and transfer learning. Given the Dirac-like nature of clicks, which is consistent across species, transfer learning techniques can be applied to existing models on new datasets with minimal additional training.
Following optimization, the model was tested on recordings from three other species: Atlantic white-sided dolphins, sperm whales, and pilot whales, to assess how the model responded to clicks from these different species. Dr. Vester provided 13 minutes and 25 seconds of unlabeled recordings from these species for testing purposes.
Two experiments were conducted using the provided recordings. The first experiment involved applying the killer whale-optimized model without any adjustments to observe the results. The second experiment included adjustments to the confidence threshold and post-processing in an attempt to improve the results without retraining the model. Retraining was not feasible due to the lack of labeled data, and no training set was generated.
The results of these experiments are shown in Figure \ref{fig:OtherAnimal}. However, these results are more anecdotal, as no annotation labels were available for comparison.

\begin{figure}[!htb] \centering
  \includegraphics[width=0.9\linewidth]{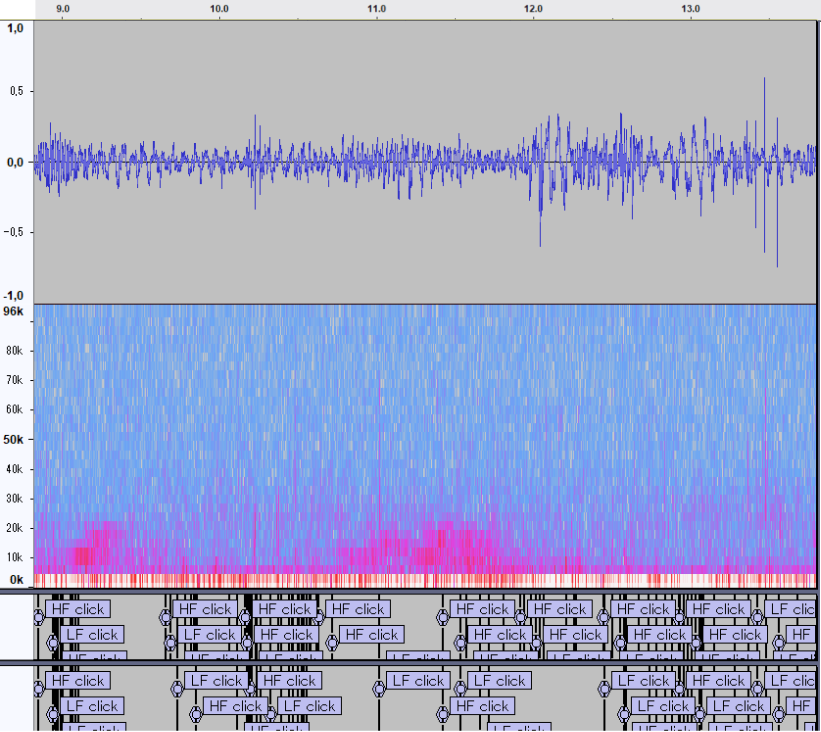}
  \caption{Extract of the Atlantic white sided Dolphins track in Audacity. The first text labels show the results of the unoptimized approach. The second text label shows the optimized approach.}
\label{fig:OtherAnimal-Dolphin}
\end{figure}
\begin{figure}[!htb] \centering
  \includegraphics[width=0.9\linewidth]{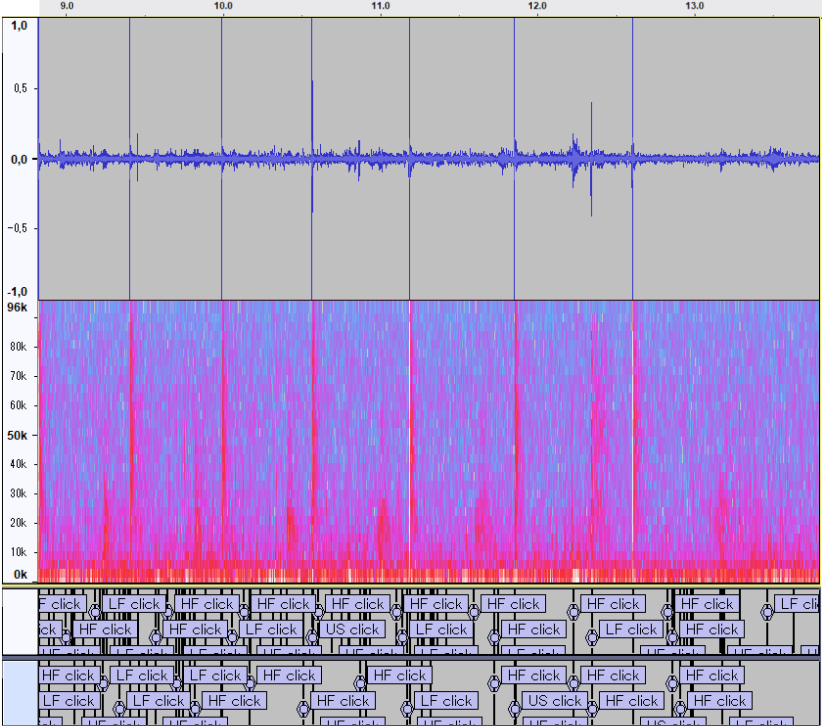}
  \caption{Extract of the pilot whale track in Audacity. The first text labels show the results of the unoptimized approach. The second text label shows the optimized approach.}
\label{fig:OtherAnimal-Pilot}
\end{figure}

Overall the unoptimized network had difficulties with the new data as the new SNR environment would trigger the YOLO threshold and, as such, the network was producing too many false positive cases. The optimized network showed visible improvements, but without proper labels, no mathematical comparisons were made. 
Finally, another exciting possibility is the possibility of investigating the phonetic structure of animal calls. The YOLO model's current usage as a click detector could be repurposed to be used as a detector for phonetic-like structures within animal calls and whistles, instead of finding Dirac-like pulsed impulses. A future tool could potentially identify subtle variations in calls that are indicative of different behaviors or intentions, further expanding its use in animal behavior studies. 

\section{Summary} 
The detection and annotation of clicks and echoes are crucial for understanding the role of clicks in communication during social interactions. However, manual annotation of large datasets is time-consuming and impractical, necessitating the development of automated solutions. While existing methods exist, many threshold-based approaches struggle in low signal-to-noise ratio (SNR) environments or fail to adequately differentiate clicks and echoes without human intervention. Given that clicks and echoes are Dirac-like pulsed signals, alternative audio representations to the spectrogram and waveform, such as continuous wavelet transforms, can convert audio into scalograms, allowing for finer-grained signal analysis. These spectrogram, scalogram and waveform representations can be encoded as grayscale images, with the option to map them to the RGB channels of an image for further processing.
Building on the limitations of the ANIMAL-SPOT model, which was too restrictive for individual event detections, the CLICK-SPOT toolchain utilized YOLO for event detection, followed by FOD post-processing to refine bounding box outputs. However, YOLO alone lacked the ability to distinguish between clicks and echoes due to insufficient contextual information in the input windows. To address this, a random forest decision tree was implemented to incorporate contextual features such as interarrival times and energy differences between peaks and means, enabling effective click-echo differentiation.
The model was trained on a dataset of 32,926 input windows and evaluated on a 2-minute, 23-second annotated recording containing 5,322 events. Experimental results indicated that while the ANIMAL-SPOT model did not perform well for this task due to technical limitations, the SCWTSPEC image representation demonstrated potential for expanding ANIMAL-SPOT’s applicability to other tasks. The YOLO network was effective for event detection, achieving an accuracy of 86.32\% with FOD post-processing. By integrating the random forest classifier to reintroduce contextual information, the CLICK-SPOT toolchain reached a click label accuracy of 95.93\% and an overall click detection accuracy of 82.56\%, outperforming other methods tested, such as PAMGuard (39.7\%), FOD-only (53.1\%), and ANIMAL-SPOT (63.9\%).
This final iteration of CLICK-SPOT shows considerable promise as an addition to bioacoustics toolkits. Future improvements in inference time could make it suitable for fieldwork as a real-time detector or validation tool. Additionally, the model could be adapted for detecting other animal vocalizations or phonetic structures in various types of acoustic data.

\section{Acknowledgments}
As the author of this thesis, I’m extremely grateful to Dr. Heike Vester and her team of biologists for providing the training data and hand labeled material. Without their dedication, time and experience this thesis would not have been possible.

\bibliographystyle{plain}
\bibliography{refs.bib}

\end{document}